\documentclass[notoc]{JHEP3}

\usepackage{graphicx}
\usepackage{amsmath,amsfonts}
\usepackage{bm}

\usepackage{color}
\let\normalcolor\relax 
\usepackage{subfigure}
\usepackage{multirow}
\usepackage{lscape}
\usepackage{rotating}

\usepackage[normalem]{ulem}

\def\hermesauthor[#1]#2{{#2}$^{\, #1}$}
\def\hermesinstitute[#1]#2{$^{#1\,}$ {#2}\\}

\def\nowat[#1]#2{\(^,\)\footnote[#1]{#2}}

\def\desy{{\sc Desy}}

\def\pythia6{{\sc Pythia6}}
\def\geant3{{\sc Geant3}}
\definecolor{Red}{rgb}{  0.6,0.0,0.0}
\definecolor{Green}{rgb}{0.0,0.6,0.0}
\definecolor{Blue}{rgb}{ 0.0,0.0,0.6}
\definecolor{Cyan}{rgb}{ 0.0,0.5,0.5}

\def\rmBH{\mathrm{BH}}
\def\rmDVCS{\mathrm{\text{DVCS}}}
\def\rmI{\mathrm{I}}

\def\Lumi{\mathcal{L}\,}

\def\tt{\bm{\eta}}

\def\AC{A_\mathrm{C}}
\def\ALUDVCS{A_{\mathrm{LU},\rmDVCS}}
\def\ALUI{A_{\mathrm{LU},\rmI}}
\def\CalAC{\mathcal{A}_\mathrm{C}}
\def\CalALUDVCS{\mathcal{A}_{\mathrm{LU}}^{\rmDVCS}}
\def\CalALUI{\mathcal{A}_{\mathrm{LU}}^{\rmI}}
\def\CalALU{\mathcal{A}_{\mathrm{LU}}}

\def\intN{{\mathcal N}}

\title{Separation of contributions from deeply virtual Compton scattering and its 
interference with the Bethe--Heitler process in measurements on a hydrogen target}

\author{The HERMES Collaboration}

\abstract{
Hard exclusive leptoproduction of real photons 
from an unpolarized proton target is studied
in an effort to elucidate generalized parton distributions.   
The data accumulated during the years 1996--2005 with the HERMES
spectrometer are analyzed to yield asymmetries 
with respect to the combined dependence of the cross section on
beam helicity and charge,
thereby revealing previously unseparated contributions from
deeply virtual Compton scattering and its interference with the Bethe--Heitler process.
The integrated luminosity is sufficient to show correlated dependences 
on two kinematic variables, and provides
the most precise determination of the dependence on only
the beam charge.
}

\keywords{Lepton-Nucleon Scattering}

\preprint{\today; version for submission}

\begin{document}

\section*{The HERMES Collaboration}

{%
\begin{flushleft} 
\bf
\hermesauthor[12,15]{A.~Airapetian},
\hermesauthor[26]{N.~Akopov},
\hermesauthor[5]{Z.~Akopov},
\hermesauthor[6]{E.C.~Aschenauer}\nowat[1]{Now at: Brookhaven National Laboratory, Upton, New York 11772-5000, USA},
\hermesauthor[25]{W.~Augustyniak},
\hermesauthor[26]{R.~Avakian},
\hermesauthor[26]{A.~Avetissian},
\hermesauthor[5]{E.~Avetisyan},
\hermesauthor[15]{B.~Ball},
\hermesauthor[18]{S.~Belostotski},
\hermesauthor[17,24]{H.P.~Blok},
\hermesauthor[5]{A.~Borissov},
\hermesauthor[13]{J.~Bowles},
\hermesauthor[19]{V.~Bryzgalov},
\hermesauthor[13]{J.~Burns},
\hermesauthor[10]{G.P.~Capitani},
\hermesauthor[21]{E.~Cisbani},
\hermesauthor[9]{G.~Ciullo},
\hermesauthor[9]{M.~Contalbrigo},
\hermesauthor[9]{P.F.~Dalpiaz},
\hermesauthor[5,15]{W.~Deconinck}\nowat[2]{Now at: Massachusetts Institute of Technology, Cambridge, Massachusetts 02139, USA},
\hermesauthor[2]{R.~De~Leo},
\hermesauthor[15,5]{L.~De~Nardo},
\hermesauthor[10]{E.~De~Sanctis},
\hermesauthor[14,8]{M.~Diefenthaler},
\hermesauthor[10]{P.~Di~Nezza},
\hermesauthor[12]{M.~D\"uren},
\hermesauthor[12]{M.~Ehrenfried}\nowat[3]{Now at: Siemens AG Molecular Imaging, 91052 Erlangen, Germany},
\hermesauthor[26]{G.~Elbakian},
\hermesauthor[4]{F.~Ellinghaus}\nowat[4]{Now at: Institut f\"ur Physik, Universit\"at Mainz, 55128 Mainz, Germany},
\hermesauthor[6]{R.~Fabbri},
\hermesauthor[10]{A.~Fantoni},
\hermesauthor[22]{L.~Felawka},
\hermesauthor[21]{S.~Frullani},
\hermesauthor[6]{D.~Gabbert},
\hermesauthor[19]{G.~Gapienko},
\hermesauthor[19]{V.~Gapienko},
\hermesauthor[21]{F.~Garibaldi},
\hermesauthor[5,18,22]{G.~Gavrilov},
\hermesauthor[26]{V.~Gharibyan},
\hermesauthor[5,9]{F.~Giordano},
\hermesauthor[15]{S.~Gliske},
\hermesauthor[10]{C.~Hadjidakis}\nowat[5]{Now at: IPN (UMR 8608) CNRS/IN2P3 - Universit\'e\ Paris-Sud, 91406 Orsay, France},
\hermesauthor[5]{M.~Hartig}\nowat[6]{Now at: Institut f\"ur Kernphysik, Universit\"at Frankfurt a.M., 60438 Frankfurt a.M., Germany},
\hermesauthor[10]{D.~Hasch},
\hermesauthor[13]{G.~Hill},
\hermesauthor[6]{A.~Hillenbrand},
\hermesauthor[13]{M.~Hoek},
\hermesauthor[5]{Y.~Holler},
\hermesauthor[6]{I.~Hristova},
\hermesauthor[23]{Y.~Imazu},
\hermesauthor[19]{A.~Ivanilov},
\hermesauthor[1]{H.E.~Jackson},
\hermesauthor[11]{H.S.~Jo},
\hermesauthor[14,11]{S.~Joosten},
\hermesauthor[13]{R.~Kaiser},
\hermesauthor[26]{G.~Karyan},
\hermesauthor[13,12]{T.~Keri},
\hermesauthor[4]{E.~Kinney},
\hermesauthor[18]{A.~Kisselev},
\hermesauthor[23]{N.~Kobayashi},
\hermesauthor[19]{V.~Korotkov},
\hermesauthor[16]{V.~Kozlov},
\hermesauthor[18]{P.~Kravchenko},
\hermesauthor[2]{L.~Lagamba},
\hermesauthor[14]{R.~Lamb},
\hermesauthor[17]{L.~Lapik\'as},
\hermesauthor[13]{I.~Lehmann},
\hermesauthor[9]{P.~Lenisa},
\hermesauthor[11]{A.~L\'opez~Ruiz},
\hermesauthor[15]{W.~Lorenzon},
\hermesauthor[6]{X.-G.~Lu},
\hermesauthor[23]{X.-R.~Lu}\nowat[7]{Now at: Graduate University of Chinese Academy of Sciences, Beijing 100049, China},
\hermesauthor[3]{B.-Q.~Ma},
\hermesauthor[13]{D.~Mahon},
\hermesauthor[14]{N.C.R.~Makins},
\hermesauthor[18]{S.I.~Manaenkov},
\hermesauthor[3]{Y.~Mao},
\hermesauthor[25]{B.~Marianski},
\hermesauthor[4]{A.~Martinez de la Ossa},
\hermesauthor[26]{H.~Marukyan},
\hermesauthor[22]{C.A.~Miller},
\hermesauthor[23]{Y.~Miyachi},
\hermesauthor[26]{A.~Movsisyan},
\hermesauthor[10]{V.~Muccifora},
\hermesauthor[13]{M.~Murray},
\hermesauthor[5,8]{A.~Mussgiller},
\hermesauthor[2]{E.~Nappi},
\hermesauthor[18]{Y.~Naryshkin},
\hermesauthor[8]{A.~Nass},
\hermesauthor[6]{M.~Negodaev},
\hermesauthor[6]{W.-D.~Nowak},
\hermesauthor[9]{L.L.~Pappalardo},
\hermesauthor[12]{R.~Perez-Benito},
\hermesauthor[8]{N.~Pickert}\nowat[8]{Now at: Siemens AG, 91301 Forchheim, Germany},
\hermesauthor[8]{M.~Raithel},
\hermesauthor[1]{P.E.~Reimer},
\hermesauthor[10]{A.R.~Reolon},
\hermesauthor[6]{C.~Riedl},
\hermesauthor[8]{K.~Rith},
\hermesauthor[13]{G.~Rosner},
\hermesauthor[5]{A.~Rostomyan},
\hermesauthor[14]{J.~Rubin},
\hermesauthor[11]{D.~Ryckbosch},
\hermesauthor[19]{Y.~Salomatin},
\hermesauthor[20]{F.~Sanftl},
\hermesauthor[20]{A.~Sch\"afer},
\hermesauthor[6,11]{G.~Schnell},
\hermesauthor[5]{K.P.~Sch\"uler},
\hermesauthor[13]{B.~Seitz},
\hermesauthor[23]{T.-A.~Shibata},
\hermesauthor[7]{V.~Shutov},
\hermesauthor[9]{M.~Stancari},
\hermesauthor[9]{M.~Statera},
\hermesauthor[8]{E.~Steffens},
\hermesauthor[17]{J.J.M.~Steijger},
\hermesauthor[12]{H.~Stenzel},
\hermesauthor[6]{J.~Stewart}\(^,\)\footnotemark[1], 
\hermesauthor[8]{F.~Stinzing},
\hermesauthor[16]{A.~Terkulov},
\hermesauthor[25]{A.~Trzcinski},
\hermesauthor[11]{M.~Tytgat},
\hermesauthor[11]{Y.~Van~Haarlem}\nowat[9]{Now at: Carnegie Mellon University, Pittsburgh, Pennsylvania 15213, USA},
\hermesauthor[11]{C.~Van~Hulse},
\hermesauthor[18]{D.~Veretennikov},
\hermesauthor[18]{V.~Vikhrov},
\hermesauthor[2]{I.~Vilardi}\nowat[10]{Now at: IRCCS Multimedica Holding S.p.A., 20099 Sesto San Giovanni (MI), Italy},
\hermesauthor[8]{C.~Vogel}\nowat[11]{Now at: AREVA NP GmbH, 91058 Erlangen, Germany},
\hermesauthor[3]{S.~Wang},
\hermesauthor[6,8]{S.~Yaschenko},
\hermesauthor[5]{Z.~Ye}\nowat[12]{Now at: Fermi National Accelerator Laboratory, Batavia, Illinois 60510, USA},
\hermesauthor[22]{S.~Yen},
\hermesauthor[12]{W.~Yu},
\hermesauthor[8]{D.~Zeiler},
\hermesauthor[5]{B.~Zihlmann}\nowat[13]{Now at: Thomas Jefferson National Accelerator Facility, Newport News, Virginia 23606, USA},
\hermesauthor[25]{P.~Zupranski}
\end{flushleft} 
}
\bigskip
{\it
\begin{flushleft} 
\hermesinstitute[1]{Physics Division, Argonne National Laboratory, Argonne, Illinois 60439-4843, USA}
\hermesinstitute[2]{Istituto Nazionale di Fisica Nucleare, Sezione di Bari, 70124 Bari, Italy}
\hermesinstitute[3]{School of Physics, Peking University, Beijing 100871, China}
\hermesinstitute[4]{Nuclear Physics Laboratory, University of Colorado, Boulder, Colorado 80309-0390, USA}
\hermesinstitute[5]{DESY, 22603 Hamburg, Germany}
\hermesinstitute[6]{DESY, 15738 Zeuthen, Germany}
\hermesinstitute[7]{Joint Institute for Nuclear Research, 141980 Dubna, Russia}
\hermesinstitute[8]{Physikalisches Institut, Universit\"at Erlangen-N\"urnberg, 91058 Erlangen, Germany}
\hermesinstitute[9]{Istituto Nazionale di Fisica Nucleare, Sezione di Ferrara and Dipartimento di Fisica, Universit\`a di Ferrara, 44100 Ferrara, Italy}
\hermesinstitute[10]{Istituto Nazionale di Fisica Nucleare, Laboratori Nazionali di Frascati, 00044 Frascati, Italy}
\hermesinstitute[11]{Department of Subatomic and Radiation Physics, University of Gent, 9000 Gent, Belgium}
\hermesinstitute[12]{Physikalisches Institut, Universit\"at Gie{\ss}en, 35392 Gie{\ss}en, Germany}
\hermesinstitute[13]{Department of Physics and Astronomy, University of Glasgow, Glasgow G12 8QQ, United Kingdom}
\hermesinstitute[14]{Department of Physics, University of Illinois, Urbana, Illinois 61801-3080, USA}
\hermesinstitute[15]{Randall Laboratory of Physics, University of Michigan, Ann Arbor, Michigan 48109-1040, USA }
\hermesinstitute[16]{Lebedev Physical Institute, 117924 Moscow, Russia}
\hermesinstitute[17]{National Institute for Subatomic Physics (Nikhef), 1009 DB Amsterdam, The Netherlands}
\hermesinstitute[18]{Petersburg Nuclear Physics Institute, Gatchina, Leningrad region 188300, Russia}
\hermesinstitute[19]{Institute for High Energy Physics, Protvino, Moscow region 142281, Russia}
\hermesinstitute[20]{Institut f\"ur Theoretische Physik, Universit\"at Regensburg, 93040 Regensburg, Germany}
\hermesinstitute[21]{Istituto Nazionale di Fisica Nucleare, Sezione Roma 1, Gruppo Sanit\`a and Physics Laboratory, Istituto Superiore di Sanit\`a, 00161 Roma, Italy}
\hermesinstitute[22]{TRIUMF, Vancouver, British Columbia V6T 2A3, Canada}
\hermesinstitute[23]{Department of Physics, Tokyo Institute of Technology, Tokyo 152, Japan}
\hermesinstitute[24]{Department of Physics, VU University, 1081 HV Amsterdam, The Netherlands}
\hermesinstitute[25]{Andrzej Soltan Institute for Nuclear Studies, 00-689 Warsaw, Poland}
\hermesinstitute[26]{Yerevan Physics Institute, 375036 Yerevan, Armenia}
\end{flushleft} 
}

\newpage

\section{Introduction}
Parton Distribution Functions (PDFs) describe
the longitudinal-momentum
structure of the nucleon in the
interpretation of inclusive and semi-inclusive 
Deep-Inelastic Scattering (DIS).
Analogously, Generalized Parton Distributions
(GPDs)~\cite{Mue94,Ji97a,Rad97} describe the multidimensional
structure of the nucleon in the
interpretation of hard exclusive leptoproduction, most simply 
when the target is left intact.
PDFs and elastic nucleon Form Factors (FFs) are embodied in GPDs 
as their limiting cases and moments, respectively~\cite {Ji97a}.
While FFs 
and PDFs represent one-dimensional distributions,
GPDs provide correlated information on 
transverse spatial and longitudinal 
momentum distributions of partons~\cite{Bur00,Bel02b,Ral02}.
In addition, the total angular momentum carried by partons 
in the nucleon can be calculated from GPDs~\cite{Ji97a}.

\begin{figure}[h]
\subfigure[]
{\includegraphics[width=0.45\columnwidth]{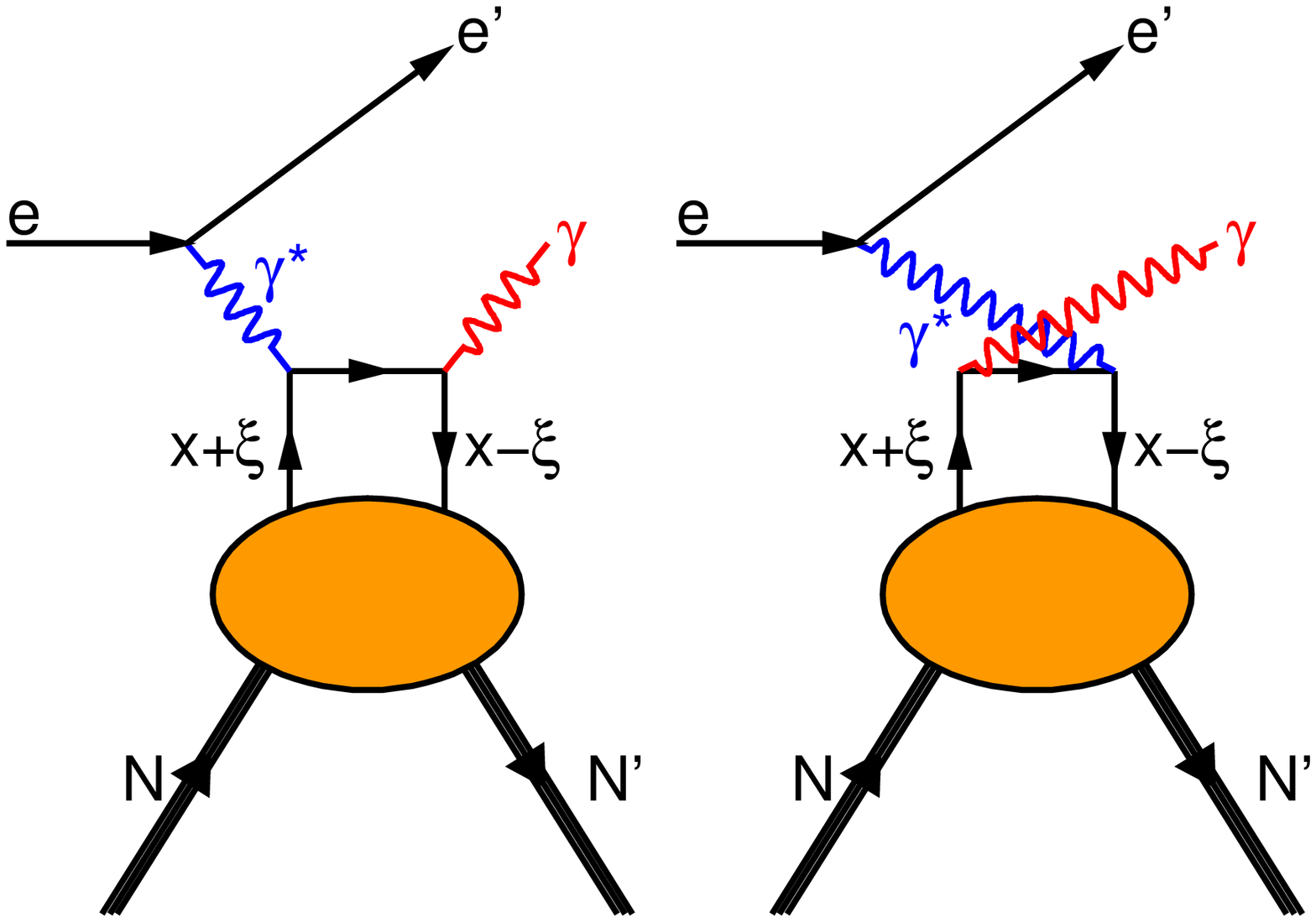}
\label{dvcs_fig}
}
\subfigure[]
{\includegraphics[width=0.45\columnwidth]{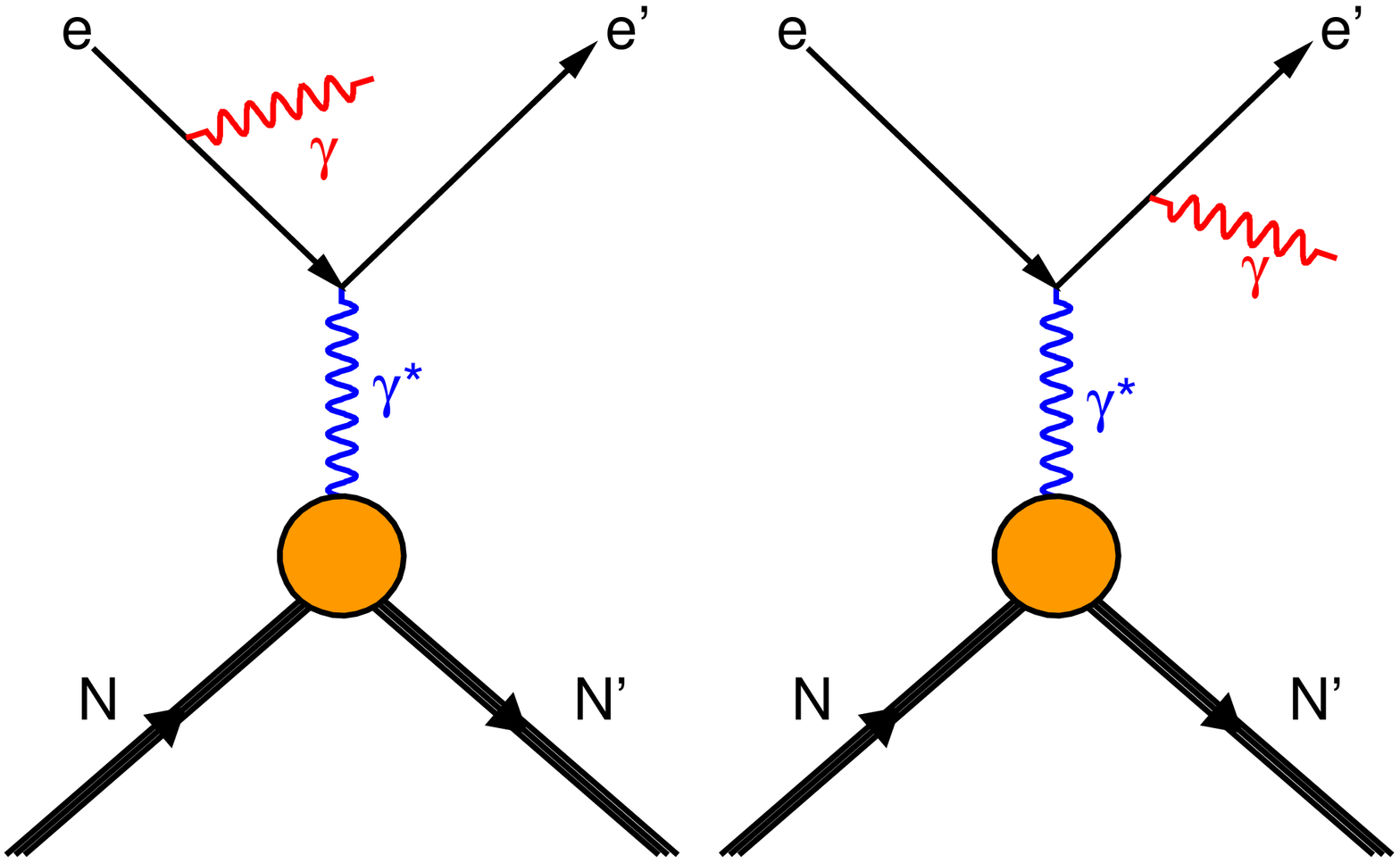}} 
\caption{Leading order diagrams for deeply virtual Compton scattering (a)
and Bethe--Heitler (b) processes.}
 \label{dvcs-bh}
\end{figure}

GPDs depend on 
the four kinematic variables $x$, $\xi$, $Q^2$, and the squared four-momentum 
transfer $t$ to the target.
In a frame where the nucleon moves with `infinite' momentum,
$x$ and $2 \xi$ are the average and difference
of the longitudinal momentum fractions of the parton in the initial and final state,
as illustrated in Fig.~\ref{dvcs_fig}.
In hard exclusive leptoproduction, 
$x$ has no direct relationship with the experimental
kinematic observable $x_{\text{B}} \equiv  Q^2 / (2 P\cdot q)$.
Here, $P$ is the four momentum of the target nucleon, $q$ is the difference between the four
momenta of the incident and scattered lepton, and  $Q^2 \equiv -q^2$.
The skewness parameter $\xi$ is related to $x_{\text{B}}$,
as $\xi \approx x_{\text{B}}/ (2-x_{\text{B}})$
in leading order Quantum Electrodynamics (QED) and in 
the kinematic limit of large $Q^2$ 
with $x_{\text{B}}$ and $t$
fixed (generalized Bjorken limit). In addition, like PDFs, GPDs are subject to Quantum Chromodynamics (QCD)
evolution with $Q^2$, which 
has been calculated perturbatively to leading order~\cite {Mue94,Ji97a,Rad97,Blu99}
and next-to-leading order~\cite {Bel99,Bel00a,Bel00b}
in the strong coupling constant $\alpha_{\text{s}}$.

Deeply Virtual Compton Scattering (DVCS),
the hard exclusive leptoproduction of real photons, e.g.,
$e^{\pm} \,  p \rightarrow e^{\pm} \,  p \, \gamma$, 
has the simplest theoretical interpretation
in terms of GPDs
among the presently experimentally feasible hard exclusive reactions.
DVCS amplitudes can be measured through the interference
between the DVCS and Bethe--Heitler (BH) processes, 
in which the photon is radiated from a parton in the former and
from the lepton in the latter (see Fig.~\ref{dvcs-bh}).
These processes have an identical final state. 
Hence 
their amplitudes $\tau_{\text{DVCS}}$ and $\tau_{\text{BH}}$ add coherently, resulting in an
interference term `$\,$I$\,$' in the cross section 
for exclusive leptoproduction of real photons.
For an unpolarized proton target,
the cross section 
can be written as~\cite {Die97,Bel02a}
\begin{align} \label {total_gamma_xsect}
\frac{\mathrm{d}\sigma}{\mathrm{d}x_{\text{B}} \, \mathrm{d}Q^2 \, \mathrm{d}|t| \, \mathrm{d}\phi} = 
\frac{x_{\text{B}} \, e^6}{32 \, (2 \pi)^4 \, Q^4 \,\sqrt{1 + \epsilon^2} }
\left[
\left| \tau_{\text{BH}} \right|^2 + 
\left| \tau_{\text{DVCS}} \right|^2 + \overbrace{
\tau_{\text{DVCS}} \, \tau_{\text{BH}}^* + \tau_{\text{DVCS}}^* \, \tau_{\text{BH}}}^{\text{I}} 
\right],
\end{align}
where 
$e$ represents the elementary charge and
$\epsilon \equiv 2 x_{\text{B}} M_p/Q$, with $M_p$ the proton mass.
The azimuthal angle $\phi$ is defined as the angle between the lepton
scattering plane and
the photon production plane spanned by the trajectories of the virtual and real photons.

The three contributions entering the
photon production cross section 
can be expanded in Fourier series in $\phi$.
For an unpolarized proton target, they
can be written as~\cite {Bel02a}
\begin{align}
|\tau_{\text{BH}}|^2 &= \frac{K_\rmBH}
{\mathcal{P}_1(\phi) \, \mathcal{P}_2(\phi)} 
\left\{ \sum_{n=0}^2 c_n^{\text{BH}} \cos(n \phi) \right\}, \quad \text {with} \quad K_\rmBH = \frac{1}{x_{\text{B}}^2 \, t \, 
(1 + \epsilon^2)^2}, \label {bh_squared} \\
|\tau_{\text{DVCS}}|^2 &= \frac{1}{Q^2}
\left\{ \sum_{n=0}^2 c_n^{\text{DVCS}} \cos(n \phi)
+ \lambda \, s_1^{\text{DVCS}} \sin \phi \right\}, \label {dvcs_squared} \\
{\text{I}} &= \frac{- e_{\ell} \, K_\rmI}{\mathcal{P}_1(\phi) \, \mathcal{P}_2(\phi)}
\left\{\sum_{n=0}^3 c_n^{\text{I}} \cos(n \phi) +
\sum_{n=1}^2 \lambda \, s_n^{\text{I}} \sin(n \phi)\right\}, \quad \text {with} \quad K_\rmI = 
\frac{1}{x_{\text{B}} \, y \, t}.  \label {I}
\end{align}
Here, $y$ is the fraction of the incident lepton energy carried by the virtual photon 
in the target rest frame, and 
$\lambda$ and $e_{\ell}$ represent respectively the beam helicity and beam charge in units
of the elementary charge. 
The  Fourier coefficients $c_n^{\text{BH}}$ and lepton 
propagators $\mathcal{P}_1(\phi)$, $\mathcal{P}_2(\phi)$ of
the BH term can be calculated within the framework of QED from 
the kinematic variables
and the Dirac and Pauli form factors $F_1$ and $F_2$ of the nucleon.

The Fourier coefficients of the interference term (Eq.~\ref{I}) are of greatest interest
since they ultimately depend on a linear combination of GPDs,
while the 
coefficients of the squared DVCS term (Eq.~\ref{dvcs_squared}) are bilinear in GPDs. 
The coefficients 
\begin{eqnarray}
s_{1}^{\rmI} &=& ~8k \, \lambda \, y (2-y) ~ \mathrm{Im}M^{1,1},  \label{s1I} \\
c_{1}^{\rmI} &=& ~8k (2-2y+y^2) ~ \mathrm{Re}M^{1,1}, \label{c1I}
\end{eqnarray}
are respectively proportional to the imaginary and real parts of $M^{1,1}$, the leading-twist
(twist-2) photon-helicity-conserving amplitude of the DVCS process.
Here, $M^{\mu,\mu^\prime}$ denotes the helicity amplitude with 
virtual (real) photon
helicity $\mu ~(\mu^\prime)$, following the notation of
Ref.~\cite {Die97}. The kinematic factor $k \propto \sqrt{-t}/Q$ 
originates from the BH propagators. Note that
the sign in Eq.~\ref{c1I} differs from that in Ref.~\cite{Bel02a}
due to the different definition of the azimuthal angle: $\phi=\pi-\phi_{{\mbox{\cite{Bel02a}}}}$.
The amplitude $M^{1,1}$ is given by a linear combination of the Compton Form Factors
(CFFs) $\mathcal{H}$, $\mathcal{\widetilde{H}}$ and $\mathcal{E}$:
\begin{equation}
M^{1,1} = F_1(t) \, \mathcal{H}(\xi,t,Q^2) + \frac{x_{\text{B}}}{2-x_{\text{B}}}
\big( F_1(t)+F_2(t) \big ) \, \mathcal{ \widetilde{H}}(\xi,t,Q^2)-\frac{t}{4M_p^2}F_2(t) \, \mathcal{E}(\xi,t,Q^2).
\label{m11}
\end{equation}
The CFFs
are convolutions of the respective twist-2 GPDs $H$, $\widetilde{H}$ or $E$,
with perturbatively calculable hard-scattering amplitudes. 
These amplitudes have been calculated to next-to-leading order in
$\alpha_{\text{s}}$~\cite {Bel98,Ji98a,Man98b}. 
The contributions from the CFFs $\mathcal{\widetilde{H}}$ and $\mathcal{E}$ 
to the amplitude $M^{1,1}$ are kinematically
suppressed compared to that from the CFF $\mathcal{H}$
at small values of $x_{\text{B}}$ and $t$, respectively.

In addition to $s_1^{\text{I}}$ and $c_1^{\text{I}}$, the only other Fourier coefficients related to 
quark-helicity-conserving twist-2 GPDs are $c_{0}^{\text{I}}$ and $c_0^{\text{DVCS}}$.
The
coefficient $c_0^{\text{I}}$ is also related to $M^{1,1}$.
Considering only the dominant CFF $\mathcal{H}$, $c_0^{\text{I}}$
is directly proportional to $c_1^{\text{I}}$ via the factor $k$ defined above:
\begin{equation}
c_{0}^{\rmI} \propto - k \, c_{1}^{\rmI}.
\label{r}
\end{equation}

The coefficients $s_{2}^{\text{I}}$, $c_2^{\text{I}}$, $s_{1}^{\text{DVCS}}$ and $c_1^{\text{DVCS}}$
are related to twist-3 GPDs.
The coefficient $s_{2}^{\text{I}}$ ($c_2^{\text{I}}$)
is proportional to the imaginary (real) part of the helicity non-conserving
amplitudes $M^{0,\pm 1}$, corresponding to the virtual photon being longitudinal. 
Conservation of angular momentum is ensured by either the exchange of an additional
gluon (genuine or dynamic twist-3) or by the fact that quarks can carry non-zero 
orbital angular momentum along the collision axis (kinematically suppressed by the same order
in $1/Q$), which is possible
due to the transverse momentum involved. 
The part of the twist-3 GPD associated with the latter picture
can be related to the twist-2 quark GPDs using 
the Wandzura--Wilczek (WW) approximation~\cite{ww77} 
and thus is also known as the WW part of the twist-3 contribution~\cite{Bel00}.

The Fourier coefficient $c_{3}^{\text{I}}$ is proportional to the real part of the 
amplitudes $M^{1,-1}$ and $M^{-1,1}$,
which 
do not conserve photon helicity, 
i.e., both photons are transverse and they have opposite helicity.
The induced two units of angular momentum can be accommodated
by gluon helicity-flip. 
Gluon helicity-flip GPDs
do not mix with quark GPDs via $Q^2$ evolution and thus probe
the gluonic properties of the nucleon~\cite{Bel00b}.
They appear at leading twist, but are suppressed by a factor $\alpha_s/\pi$.
In addition, as in the case of the coefficients discussed above that are kinematically 
suppressed by $1/Q$,
it is possible that the two participating quarks complete the conservation of angular momentum 
if they carry orbital angular momentum. As they have to account for
two units of angular momentum instead of one as above, 
this process appears at twist-4. The associated twist-4 GPD was found to be 
calculable in terms of twist-2 quark GPDs using the WW
approximation~\cite {Kiv01b}. Similarly, the Fourier coefficient $c_{2}^{\text{DVCS}}$
arises from the twist-2 gluon helicity-flip GPDs with possible contributions
from twist-4 quark GPDs.

\section {Asymmetries}
Previous measurements with a longitudinally (L) polarized positron \{electron\}
beam by HERMES~\cite{Air01} \{CLAS~\cite{Ste01,Gir08,Gav08}\} on an unpolarized (U) proton target
provided access to a combination of
$s_1^\rmI$ and $s_1^{\text{DVCS}}$ via the 
{\it single-charge} beam-helicity asymmetry, also
denoted as the Beam-Spin Asymmetry (BSA):
\begin{align} 
& \CalALU (\phi,e_{\ell}) \equiv
\frac { d\sigma^{\to} -  d\sigma^{\gets}}
{ d\sigma^{\to} + d\sigma^{\gets}} \nonumber \\
& = \frac{ - e_{\ell} \frac{K_\rmI}{{\mathcal P}_1(\phi){\mathcal
      P}_2(\phi)} \left[ \displaystyle\sum_{n=1}^2 s_{n}^\rmI \sin(n\phi) \right] 
+ \frac{1}{Q^2} s_1^{\text{DVCS}} \sin \phi}{\frac{1}{{\mathcal P}_1(\phi){\mathcal P}_2(\phi)} \left[K_\rmBH
\displaystyle\sum_{n=0}^2 c_{n}^\rmBH \cos(n\phi) - e_{\ell} K_\rmI 
 \displaystyle\sum_{n=0}^3 c_{n}^\rmI \cos(n\phi) \right] + \frac{1}{Q^2}
\displaystyle\sum_{n=0}^2 c_{n}^\rmDVCS \cos(n\phi)}.
\label{bsa_old}
\end{align}
Here, $\sigma^{\to}$ ($\sigma^{\gets}$) 
denotes the cross section for a beam with positive (negative) helicity.
Predominant $\sin \phi$ dependences with opposite sign have been
observed at the two experiments,
indicating the dominance of the interference term involving $e_{\ell} \cdot s_1^\rmI$.
However, quantitative access to $s_1^\rmI$ is complicated by the presence 
of $s_1^{\text{DVCS}}$, which is a higher twist-contribution 
but possibly significant, and by the presence of $c_1^{\text{I}}$ and $c_0^{\text{I}}$,
i.e., the other Fourier coefficients of interest appearing at 
leading twist (see Eqs. \ref{c1I} and \ref{r}).

This entanglement can be avoided by defining the 
{\it charge-difference} beam-helicity asymmetry~\cite{Ell05}: 
\begin{align}
\CalALUI (\phi) &\equiv \frac{(d \sigma^{+\to} - d \sigma^{+\gets})
  - (d \sigma^{-\to} - d \sigma^{-\gets})}
{(d \sigma^{+\to} + d \sigma^{+\gets})
  + (d \sigma^{-\to} + d \sigma^{-\gets})}
\nonumber \\
&= \frac{- \frac{K_\rmI}{{\mathcal P}_1(\phi){\mathcal
      P}_2(\phi)} \left[ \displaystyle\sum_{n=1}^2 s_{n}^\rmI\sin(n\phi)
  \right]}{\frac{K_\rmBH}{{\mathcal P}_1(\phi){\mathcal P}_2(\phi)}\displaystyle\sum_{n=0}^2
  c_{n}^\rmBH \cos(n\phi)+ \frac{1}{Q^2} \displaystyle\sum_{n=0}^2 c_{n}^\rmDVCS
  \cos(n\phi)},
\label{eq:ALUI}
\end{align}
where the additional $+(-)$ superscript on the cross-sections
denotes the charge of the lepton beam.
This asymmetry 
has the important advantages that the $\sin \phi$ dependence in the numerator
stems solely from the interference term, as the (higher-twist) $\sin \phi$ dependence of the
squared DVCS term cancels, and the denominator no longer contains the leading terms 
$c_1^{\text{I}}$ and $c_0^{\text{I}}$.
Therefore it gives direct access to linear combinations of GPDs, while 
another {\it charge-averaged} asymmetry related to the squared DVCS term provides access to bilinear 
combinations of GPDs:
\begin{align}
\CalALUDVCS (\phi) &\equiv \frac{(d \sigma^{+\to} - d \sigma^{+\gets})
  + (d \sigma^{-\to} - d \sigma^{-\gets})}
{(d \sigma^{+\to} + d \sigma^{+\gets})
  + (d \sigma^{-\to} + d \sigma^{-\gets})}
\nonumber \\
&= \frac{\frac{1}{Q^2} \,           s_{1}^\rmDVCS\sin \phi}
{\frac{K_\rmBH}{{\mathcal P}_1(\phi){\mathcal P}_2(\phi)}\displaystyle\sum_{n=0}^2
  c_{n}^\rmBH \cos(n\phi)+ \frac{1}{Q^2} \displaystyle\sum_{n=0}^2 c_{n}^\rmDVCS
  \cos(n\phi)}.
\label{eq:ALUDVCS} 
\end{align}
The previously extracted~\cite{Air06, Air08} Beam-Charge Asymmetry (BCA) 
\begin{align}
\CalAC(\phi) &\equiv \frac{d\sigma^+ - d\sigma^-} {d\sigma^+ + d\sigma^-} 
= \frac{-\frac{K_\rmI}{{\mathcal P}_1(\phi){\mathcal P}_2(\phi)} \displaystyle\sum_{n=0}^3
  c_{n}^I \cos(n\phi)}{\frac{K_\rmBH}{{\mathcal P}_1(\phi){\mathcal
      P}_2(\phi)}\displaystyle\sum_{n=0}^2 c_{n}^\rmBH \cos(n\phi)+ \frac{1}{Q^2} \displaystyle\sum_{n=0}^2
  c_{n}^\rmDVCS \cos(n\phi)}
\label{eq:AC}
\end{align}
provides access to the real part of the DVCS amplitude via $c_{n}^\rmI$.

\section{Event selection}
The data were collected during the years 1996--2005 with the HERMES
spectrometer~\cite {Ack98} using the longitudinally polarized 27.6 GeV 
electron and positron beams provided by the HERA accelerator facility at DESY.
The hydrogen gas target was either unpolarized, longitudinally or
transversely nuclear-polarized. However, the time averaged polarization of the
polarized targets was negligible, while the rapid ($60-180\,$s)
reversal of the polarization direction minimized polarization bias due 
to detector effects.
The polarization direction of the beam was reversed about every two months.
The integrated luminosity for the electron (positron) data sample 
corresponds to about $106 \, \text{pb}^{-1} (292 \, \text{pb}^{-1})$ with an average 
magnitude of the beam polarization of 30.0$\,$\% (49.4$\,$\%). The latter has a mean
fractional systematic uncertainty of 2.8$\,$\%.

A brief description of the event selection is given here. 
More details can be found in Refs.~\cite{Air06, Ell04}.
Events are selected with
exactly one photon producing an energy deposition larger than $5~\text{GeV}~(1~\text{MeV})$ 
in the calorimeter (preshower detector) and one charged track, 
identified as the scattered lepton, in the kinematic range $1~\text{GeV}^2 < Q^2 < 10~\text{GeV}^2,
0.03 < x_B < 0.35$, $W > 3$~GeV and $\nu < 22~\text{GeV}$. 
Here, $W$ denotes the invariant mass of the initial photon-nucleon system and $\nu$ 
denotes the virtual-photon energy in the target rest frame.
The angle $\theta_{\gamma^{*}\gamma}$ between the real and the virtual photon is 
constrained to be between $5$ and $45 ~\text{mrad}$.
The recoiling proton is not detected. An `exclusive' sample of events 
is selected by the requirement that the  
squared missing mass $M_X^2$ of the reaction $e^{\pm} \, p \rightarrow e^{\pm} \, \gamma \, X$ 
corresponds to the squared proton mass.
The resolution in $M_X^2$ is limited by the energy resolution of the real photon
in the calorimeter. 
Correspondingly, the exclusive region is defined as $-(1.5~\text{GeV})^2 < M_X^2 < (1.7~\text{GeV})^2$, 
as determined from signal-to-background studies using a Monte Carlo (MC)
simulation. 
For elastic events (leaving the proton intact), the
kinematic relationship between the energy and direction of the real photon permits $t$
to be calculated without using the measured energy of the real photon, which is the
quantity subject to the largest uncertainty.
Thus, the value of $t$ 
is calculated as
\begin{equation} \label{tc}
t = \frac{-Q^2 - 2 \, \nu \, (\nu - \sqrt{\nu^2 + Q^2} \, \cos\theta_{\gamma^* \gamma })}
{1 + \frac{1}{M_p} \, (\nu - \sqrt{\nu^2 + Q^2} \, \cos\theta_{\gamma^* \gamma })}
\end{equation} 
for the exclusive event sample. The quantity $-t$ is
required to be smaller than $0.7~\text{GeV}^2$. 
The error caused by applying this expression to
inelastic events
is accounted for in the MC simulation that is used to calculate
the fractional contribution of background processes
per kinematic bin in $x_{\text{B}}, Q^2$, and $-t$.

\section{Extraction of asymmetry amplitudes}
The experimental yield $\intN$ can be parameterized as
\begin{align}
\intN(e_{\ell},P_{\ell},\phi)=
\Lumi (e_{\ell},P_{\ell}) \eta(e_{\ell},\phi) \sigma_{\text{UU}}(\phi) \times 
\left[ 1 + P_{\ell} \CalALUDVCS(\phi) + e_{\ell} P_{\ell} \CalALUI(\phi) 
+ e_{\ell}\CalAC(\phi) \right].
\end{align}
Here, $\Lumi$ is the integrated luminosity, $\eta$ the detection efficiency, $P_{\ell}$ the beam
polarization and $\sigma_{\text{UU}}$ 
the cross section for an unpolarized target averaged over both beam charges and helicities. 
The asymmetries $\CalALUI(\phi)$, $\CalALUDVCS(\phi)$, and $\CalAC(\phi)$ relate to the Fourier coefficients
in Eqs.~\ref{bh_squared}--\ref{I} according to Eqs.~\ref{eq:ALUI}--\ref{eq:AC}. 
They are expanded in $\phi$ as
\begin{eqnarray}
\CalALUI(\phi)& \simeq &\sum_{n=1}^2 \ALUI^{\sin(n\phi)}\sin(n\phi) \quad {+ \sum_{n=0}^1 \ALUI^{\cos(n\phi)}\cos(n\phi)},
\label{eq:asymmetry2} 
\end{eqnarray}
\begin{eqnarray}
\CalALUDVCS(\phi)& \simeq &\sum_{n=1}^2 \ALUDVCS^{\sin(n\phi)}\sin(n\phi)
 \quad {+ \sum_{n=0}^1 \ALUDVCS^{\cos(n\phi)}\cos(n\phi)},
\label{eq:asymmetry3} 
\end{eqnarray}
\begin{eqnarray}
\CalAC(\phi)& \simeq &\sum_{n=0}^3 \AC^{\cos(n\phi)}\cos(n\phi) \quad {+
\AC^{\sin \phi}\sin \phi}, \label{eq:asymmetry1} 
\end{eqnarray}
where the approximation is due to the truncation of the
in general infinite Fourier series caused by the azimuthal
dependences in the denominators of Eqs.~\ref{eq:ALUI}--\ref{eq:AC}.
The asymmetry amplitudes $\ALUI^{\sin \phi}$, $\AC^{\cos \phi }$ and 
$\AC^{\cos(0\phi)}$ relate to the twist-2 Fourier coefficients of the interference
term appearing in Eq. \ref{I} and further developed in Eqs.~\ref{s1I},
\ref{c1I} and \ref{r}, respectively. 
The asymmetry amplitudes $\ALUDVCS^{\sin \phi }, \ALUI^{\sin(2\phi)}, \AC^{\cos(2\phi)}$ and
$\AC^{\cos(3\phi)}$ are related to other
Fourier coefficients in Eqs.~\ref{dvcs_squared} and \ref{I}, which are also explained above. 
The remaining asymmetry amplitudes are expected to be zero
but were introduced to test for instrumentally induced harmonics.
(The asymmetry amplitude  $\ALUDVCS^{\sin(2\phi)}$ can, in addition, arise
through the interplay of numerator and denominator in Eq.~\ref{eq:ALUDVCS}  
if the twist-3 Fourier coefficient $s_{1}^{\text{DVCS}}$ has a sizeable value.) 
Comparison of predictions based on GPD models to data for either asymmetry amplitudes 
or the Fourier coefficients in Eqs.~\ref{dvcs_squared} and \ref{I} provides 
similar information,
as the asymmetry amplitudes relate to the corresponding Fourier coefficients 
in an approximately model-independent way.
This is due to the fact that the BH coefficients $c_n^{\text{BH}}$ and lepton propagators 
are precisely calculable, and that in the kinematic region of HERMES the contributions from
the squared DVCS term in the denominators of Eqs.~\ref{eq:ALUI}--\ref{eq:AC} are expected to be much 
smaller than the BH contributions~\cite{Kor02a}.

The extraction of the asymmetry amplitudes in each kinematic bin of $x_B$, $Q^2$ and $t$
is based on the maximum likelihood technique~\cite{Bar90}, 
providing a bin-free fit in $\phi$ to the data. 
Its application here is explained in detail in
Ref.~\cite{Air08}. In the fit, weights were introduced to account for
luminosity imbalances with respect to beam charge and polarization.
No balancing procedure was required for the target polarization, 
as the time averaged polarization was negligible
as mentioned above.

\section{Background corrections and systematic uncertainties}
Each extracted asymmetry amplitude $A$ is corrected, in each kinematic bin,
for the contribution of semi-inclusive and exclusive background, which is mostly due to
the production of $\pi^{0}$ and $\eta$ mesons.
The background-corrected amplitude is calculated as
\begin{equation}
A_{\text{corr}} = \frac{A -f_{\text{semi}} A_{\text{semi}} - f_{\text{excl}} A_{\text{excl}}}  {1 - f_{\text{semi}} - f_{\text{excl}}  }.
\end{equation}
The fraction  $f_{\text{semi}}$ of semi-inclusive background 
per bin is calculated using a MC
simulation described in detail in Ref~\cite{Air08}.
It varies between $0.6\%$ and $12.7\%$ depending on
the kinematic bin, with an average value of $3.3\%$.
Based on the model in Ref.~\cite {Van99}, the fraction $f_{\text{excl}}$ of exclusive
background is estimated 
in each kinematic bin and found to be below $0.7\%$ everywhere.
A direct search for exclusive neutral pions in the HERMES data supports this estimate~\cite{Arne}.
The semi-inclusive (exclusive) background can have a non-zero 
asymmetry $A_{\text{semi}}$ ($A_{\text{excl}}$),
as has been measured for, e.g., the semi-inclusive production of $\pi^0$ mesons,
which exhibits a sinusoidal $\phi$ dependence on the beam helicity~\cite{hermes_pi0_ssa}.
The beam-charge-dependent background asymmetry is
zero at leading order QED.
Hence
the contributions from semi-inclusive and exclusive background constitute a dilution 
of $\CalAC$ and effectively also of $\CalALUI$, as the latter involves only charge differences.
In order to correct $\CalALUDVCS$ for the semi-inclusive background, the size
of the beam-helicity asymmetry for this background is extracted from data
by reconstructing neutral pions with a large fractional 
energy $E_{\pi^{0}}/\nu > 0.8$ (see Ref.~\cite{Air08} for details). 
For the exclusive background, the asymmetry cannot be extracted
from data due to the small yield of exclusive pions. 
As the asymmetry is 
in the range [-1,1], a value of zero is assumed with a ``statistical'' uncertainty 
of $2/\sqrt{12}$, i.e., 
one standard deviation for a uniform distribution.
The statistical uncertainty on the total
correction due to the statistical uncertainties in the background fractions and asymmetries
is propagated as a contribution to the final statistical uncertainty. 
In addition, half the size of the actual
correction is assigned as systematic uncertainty 
to account for assumptions and approximations inherent in the approach.

The dominant systematic uncertainty for the amplitude $A_{\text{LU,DVCS}}^{\sin\phi}$ is
due to the background correction.
In the case of $A_{\text{LU,I}}^{\sin (n\phi)}$ and $A_{\text{C}}^{\cos (n\phi)}$, the systematic 
uncertainty is predominantly due 
to the combined contributions of possible deviations of the detector and/or the beam from their
nominal positions (`alignment'), detector acceptance including smearing, and finite bin width in
$x_{\text{B}}, t$ and $Q^2$.
The systematic uncertainty arising from these combined contributions is estimated by MC
simulations using the GPD model described in
Ref~\cite{Guz06}.
Note that a mistake has been found in this GPD
model~\cite{Guz08}. However, the model of Ref.~\cite{Guz06} described
HERMES beam-charge~\cite{Air08} and preliminary (single charge) 
beam-helicity asymmetries~\cite{Ell07} and thus is considered adequate for systematic studies.
In each bin, the systematic uncertainty is taken as the difference between
the model predictions at the mean
kinematic value of that bin and the respective amplitude extracted from the
reconstructed MC data.

Further systematic uncertainties arise from 
an observed relative shift of the squared missing-mass spectra between the
electron and positron sample, with a magnitude of approximately 0.2 GeV$^2$~\cite{Zei09}. 
The boundaries defining the exclusive sample
in the missing mass spectra were adjusted to account for this shift. One quarter
of the effect on the extracted asymmetries is applied as systematic uncertainty.
The impact
of both trigger and tracking inefficiencies was studied and found to
be negligible.
Also not included is any contribution due to additional QED vertices, as the most
significant of these has been estimated to be negligible in the case
of helicity asymmetries~\cite{Afa05}.
The systematic uncertainties are added in quadrature and given in Table~\ref{sys-sum}.
\TABLE[t]{\scriptsize
\begin{tabular}{|l|r|c|c|c|} \hline
Amplitude & $A \pm \delta (\text{stat.}) \pm \delta (\text{syst.})$ & Missing
mass shift & Background corr. & Alignment, acceptance, bin width \\ \hline \hline
$A_{\text{LU,I}}^{\sin \phi}$    & $ -0.224 \pm 0.028 \pm 0.020$ & $0.001$ & $0.005$ & $0.019$ \\ \hline
$A_{\text{LU,DVCS}}^{\sin \phi}$ & $ ~0.043 \pm 0.028 \pm 0.004$ & $0.000$ & $0.004$ & $0.000$ \\ \hline
$A_{\text{C}}^{\cos(0\phi)}$     & $ -0.020 \pm 0.006 \pm 0.008$ & $0.004$ & $0.000$ & $0.007$ \\ \hline
$A_{\text{C}}^{\cos \phi}$      & $ ~0.055 \pm 0.009 \pm 0.004$ & $0.001$ & $0.001$ & $0.003$ \\ \hline
\end{tabular}
\caption{
Overall values of the asymmetry amplitudes of greatest interest, 
at average kinematics $\langle -t \rangle = 0.12$~GeV$^2$, 
$\langle x_B \rangle= 0.09$, and $\langle Q^2 \rangle =2.37$~GeV$^2$.
The rightmost three columns present dominant contributions to the systematic uncertainties. 
Not included is a 2.8\% scale uncertainty of the beam-helicity asymmetries due to 
the beam polarization measurement.}
\label{sys-sum}
}

\section{Results}
Table~\ref{sys-sum} presents
the asymmetry amplitudes of greatest interest extracted in the entire HERMES kinematic
acceptance (``overall'' results).
The $\sin \phi$ amplitude of the beam-helicity asymmetry sensitive
to the interference term
is shown in the first row of Fig.~\ref{bsa_fig}.
\FIGURE{
\includegraphics[width=\columnwidth]{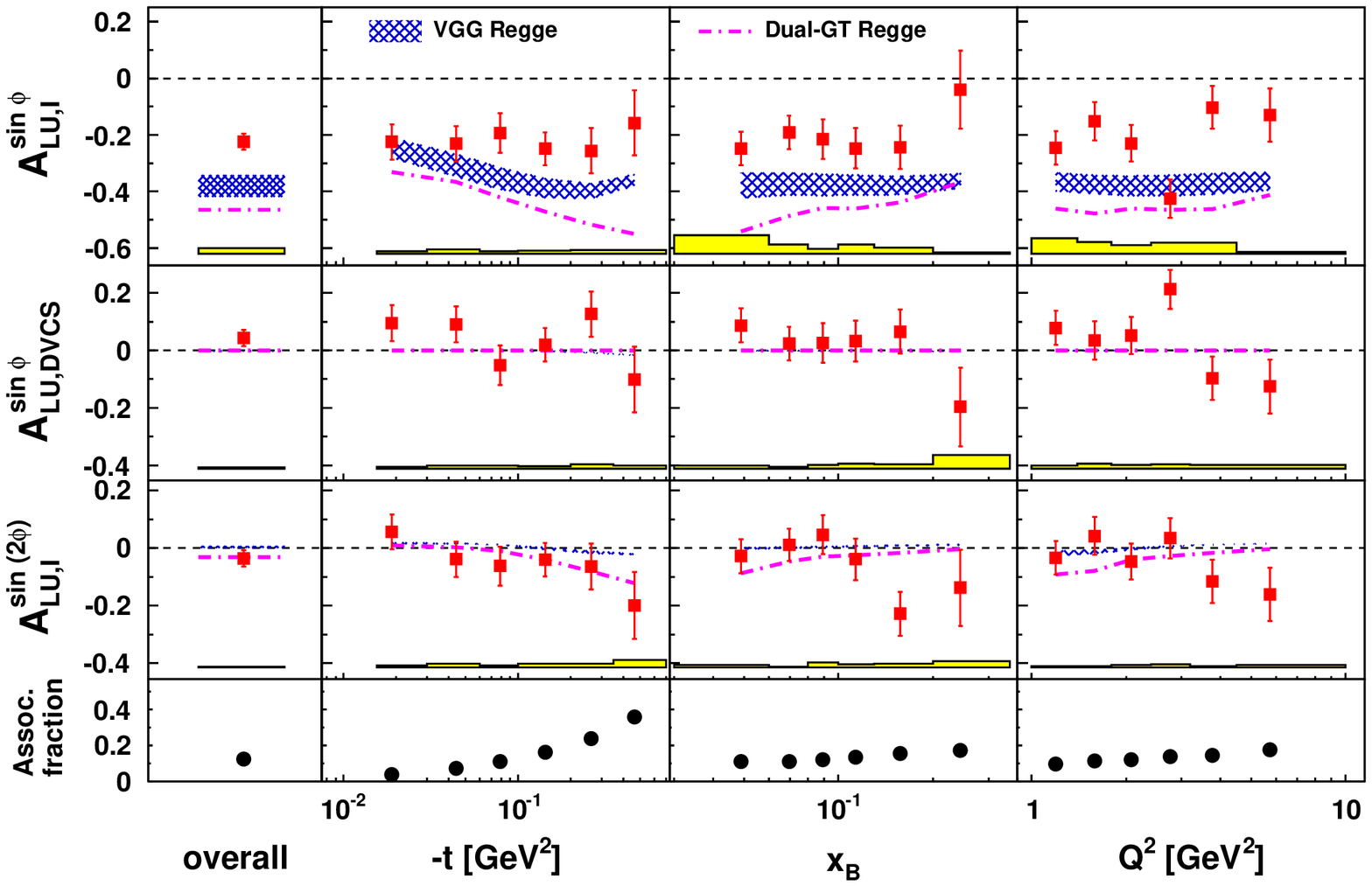}
\caption{
The first (second) row shows the $\sin \phi$ amplitude 
of the beam-helicity asymmetry $\ALUI$ ($\ALUDVCS$), which is sensitive 
to the interference term (squared DVCS term),
extracted from the 1996--2005 hydrogen data in the entire experimental acceptance, and 
as a function of $-t$, $x_{\text{B}}$, and $Q^2$. 
The third row shows the $\sin 2 \phi$ amplitude of $\ALUI$.
The error bars (bands) represent the statistical (systematic) uncertainties.
Not included is a 2.8\% scale uncertainty due to 
the beam polarization measurement.
The calculations are 
based on the recently corrected minimal implementation~\cite{Guz06,Guz08}  
of a dual-parameterization GPD model (Dual--GT) and 
on a GPD model~\cite{Van99,Goe01} based on double--distributions (VGG).
Both models use a Regge--motivated $t$-dependence. 
The band for the VGG model results from 
varying the parameters $b_{\text{val}}$ and $b_{\text{sea}}$ between unity and infinity.
The bottom row shows the fractional contribution of associated BH production 
as obtained from a MC simulation.}
 \label{bsa_fig}
}
It exhibits a large overall value of 
$\ALUI^{\sin\phi} = -0.224 \pm 0.028(\text{stat.}) \pm 0.020$(sys.),
with no significant dependence
on any of the kinematic variables $-t$, $x_{\text{B}}$, and $Q^2$.
This implies a rather strong dependence of this amplitude 
on $t$ for smaller values of $-t$, as the asymmetry amplitude 
has to vanish in the limit of vanishing $-t$ due to the vanishing factor $k$ 
in Eq.~\ref{s1I}.
(In the limit of vanishing $t$, $c_0^{\text {BH}}$ remains finite and the dependences of 
$K_{\text {BH}}$, $K_{\text I}$, ${\mathcal P}_1(\phi)$ and ${\mathcal P}_2(\phi)$ 
on $t$ cancel in Eq.~\ref{eq:ALUI}.)
The $\sin \phi$ amplitude of the beam-helicity asymmetry sensitive
to the squared DVCS term
is shown in the second row of Fig.~\ref{bsa_fig}.
It also shows no kinematic dependence, with an overall
value of $\ALUDVCS^{\sin\phi} = 0.043 \pm 0.028(\text{stat.}) \pm 0.004$(sys.).
As explained above (see Eq. \ref{bsa_old}), the beam-helicity asymmetries measured previously
with a single beam-charge are sensitive only to the combination of the results 
presented here, i.e., the single-beam-charge results are given as 
\begin{equation}
A_{\mathrm{LU}}^{\sin\phi}( e_{\ell}) \approx e_l \, \ALUI^{\sin\phi} + \ALUDVCS^{\sin\phi},
\label{sinphiapprox}
\end{equation}
if the contributions $c_n^{\text{I}}$ from the interference term in the denominator 
of Eq.~\ref{bsa_old} can be neglected. Previous HERMES measurements~\cite{Air06,Air08} 
found these contributions to be small 
compared to the remainder of the denominator, and a more precise constraint is presented below.
Using the present data, the separate analysis of the positron \{electron\} data yields 
values for $A_{\mathrm{LU}}^{\sin\phi}( e_{\ell})$ of $-0.177 \pm 0.022(\text{stat.})$ 
\{$0.255 \pm 0.051(\text{stat.})$\}, 
in agreement with $-0.181 \pm 0.046(\text{stat.})$ \{$0.267 \pm 0.065(\text{stat.})$\}
calculated from Eq.~\ref{sinphiapprox} neglecting correlations from the
commonality of the data sets.

The $\sin 2 \phi$ amplitude of $\ALUI$ is shown in the third row of Fig.~\ref{bsa_fig}.
It has an overall value consistent with zero ($-0.035 \pm 0.028 \pm 0.002$) and thus, 
as expected, is much smaller than the corresponding $\sin \phi$ amplitude.
Those asymmetry amplitudes included
in Eqs.~\ref{eq:asymmetry2}--\ref{eq:asymmetry1} as tests for instrumental effects
are found to be compatible with zero.

The data in Fig.~\ref{bsa_fig} are compared with theoretical calculations 
to leading order in QED and QCD.
In the GPD based model
(VGG) 
described in Refs.~\cite{Van99,Goe01}, 
the dependences on $\xi$ and $t$ are factorized while those on $x$ and $t$ may
be entangled when a Regge--motivated ansatz is invoked. The model is formulated as a 
double distribution~\cite{Mue94,Rad97} complemented by a D--term~\cite{Pol99}, 
where the kernel of the double distribution
contains a profile function~\cite{Rad99a,Rad99b} that determines the dependence on $\xi$, 
controlled by a parameter $b$~\cite{Mus00}.
In the limit $b \rightarrow \infty$ the GPD is independent of $\xi$. 
Note that $b_{\mathrm{val}}$ ($b_{\mathrm{sea}}$) is a free parameter 
for the valence (sea) quarks and thus can be
used as a fit parameter in the extraction of GPDs from hard-electroproduction data.

In each kinematic bin, a range of theoretical predictions was calculated~\cite{Van01}
by varying the model parameters of only the GPD $H$,
since these data are sensitive mostly to this GPD as explained above.
Variants of the model are distinguished by differences in
the $t$ dependence of the GPD $H$, for which
either a simple ansatz is used where 
the $t$ dependence factorizes from the dependence on the other kinematic variables,
or the Regge--motivated ansatz is employed.
Since the differences are found to be small for all amplitudes
shown in Fig.~\ref{bsa_fig}, 
only the results based on the latter ansatz (VGG Regge) are displayed.
The broad width of the bands is due
to the fact that the parameters $b_{\mathrm{val}}$ and $b_{\mathrm{sea}}$ 
were varied between unity and infinity, 
with the variation in $b_{\text{sea}}$ having the strongest effect.
Note that including or neglecting the D--term in the GPD model 
does not change the result since it
contributes only to the real part of the DVCS amplitude.
The other model presented here (Dual--GT) is  based on the corrected~\cite{Guz08} 
minimal implementation~\cite{Guz06} of the dual parameterization GPD model~\cite{Shu02},
in which the dependence on $\xi$ is factorized from the dependences on $x$ and $t$.
The $t$ dependence in this model is also 
Regge--motivated.
All models overestimate the magnitude of $\ALUI^{\sin\phi}$ by
approximately a factor of two. They are consistent with the observed shapes of the 
kinematic dependences on $x_B$ and $Q^2$ (but not $t$).
The part of the VGG band closest to the data, i.e.,
with the smallest absolute amplitude, 
corresponds to $b_{\text{sea}}= \infty$. 
Note that these models are for the elastic part 
of the cross section only while this measurement includes associated
production in which the nucleon in the final
state is excited to a resonant state.
In the following it is considered
whether the contribution from associated production
can account for the observed discrepancies between model predictions and data.

The bottom row of Fig.~\ref{bsa_fig} shows the estimated fractional 
contribution from associated BH production
in each kinematic bin,
calculated using MC simulations described in Ref.~\cite{Air08}.
The overall value is about 12\%.
In an attempt to estimate  $\ALUI^{\sin\phi}$   
in elastic and associated 
production separately, the strong dependence of the fractional contributions
of elastic and associated production on the missing mass value  
in the exclusive region $-(1.5~\text{GeV})^2 < M_X^2 < (1.7~\text{GeV})^2$
can be utilized.
The exclusive region can be split
in several bins, each bin with its 
background-corrected amplitude 
$A_{\text{corr}} = f_{\text{elas}} A_{\text{elas}} + f_{\text{asso}} A_{\text{asso}}$.
The fraction $f_{\text{elas}}$ ($f_{\text{asso}}$) of elastic (associated) production per
bin is taken from the MC simulation, in which the DVCS process is not implemented
because 
the BH cross section is expected to dominate
that of DVCS not only for elastic but also for associated production~\cite{Gui03}.    
Assuming that the values of $\ALUI^{\sin\phi}$ for elastic and associated production 
do not depend on $M_X^2$, 
the two unknown amplitudes $A_{\text{elas}}$ and $A_{\text{asso}}$ are extracted from the five equations 
corresponding to the five $M_X^2$--bins.
The resulting overall $\sin \phi$ amplitude from elastic production is found to be $-0.209 \pm 0.066$
and thus hardly differs from that
reported in Table~\ref{sys-sum}, while the one from associated production 
can only be constrained 
to be between $-0.68$ and $0.09$ within one standard deviation in 
the statistical uncertainty.
According to theoretical calculations~\cite{Gui03} 
a correction factor of 1.1 has to be applied to the measured 
beam-helicity asymmetry in HERMES kinematics
due to the $\Delta$ resonance region ($W < 1.35$~GeV).
For the associated DVCS amplitude these calculations are based on a model for transition GPDs, 
which are related to those on the nucleon within this model.
Thus neither the 
extracted $\sin \phi$ amplitudes $A_{\text{elas}}$ and $A_{\text{asso}}$
nor model calculations support
extreme scenarios in which the $\sin \phi$ amplitude from associated production
has an overall value of unity, 
which would be required to obtain 
a $\sin \phi$ amplitude for elastic production
as large as the value $-0.39$ or more predicted by the models shown.
Thus associated production cannot account for the overestimate of the asymmetry amplitudes 
by the models.

A promising alternative 
to comparing the data with existing models
is to use a flexible GPD parameterization and perform a global fit to all DVCS data. 
First steps in this direction have been made~\cite{Kum08,Gui08, Kum09, Mou09, Gui09},
one of which found that a preliminary HERMES result on the $\sin \phi$ amplitude
of the beam helicity asymmetry for a single charge (positron)~\cite{Ell07} can be described
by a fit to other DVCS data~\cite{Kum09}.
In order to provide additional input for future fits, in particular 
for the entangled $\xi$ and $-t$ dependences of GPDs, the amplitudes
already presented in Fig.~\ref{bsa_fig} are shown in Fig.~\ref{2dim_bsa_fig} 
\FIGURE{
\includegraphics[width=\columnwidth]{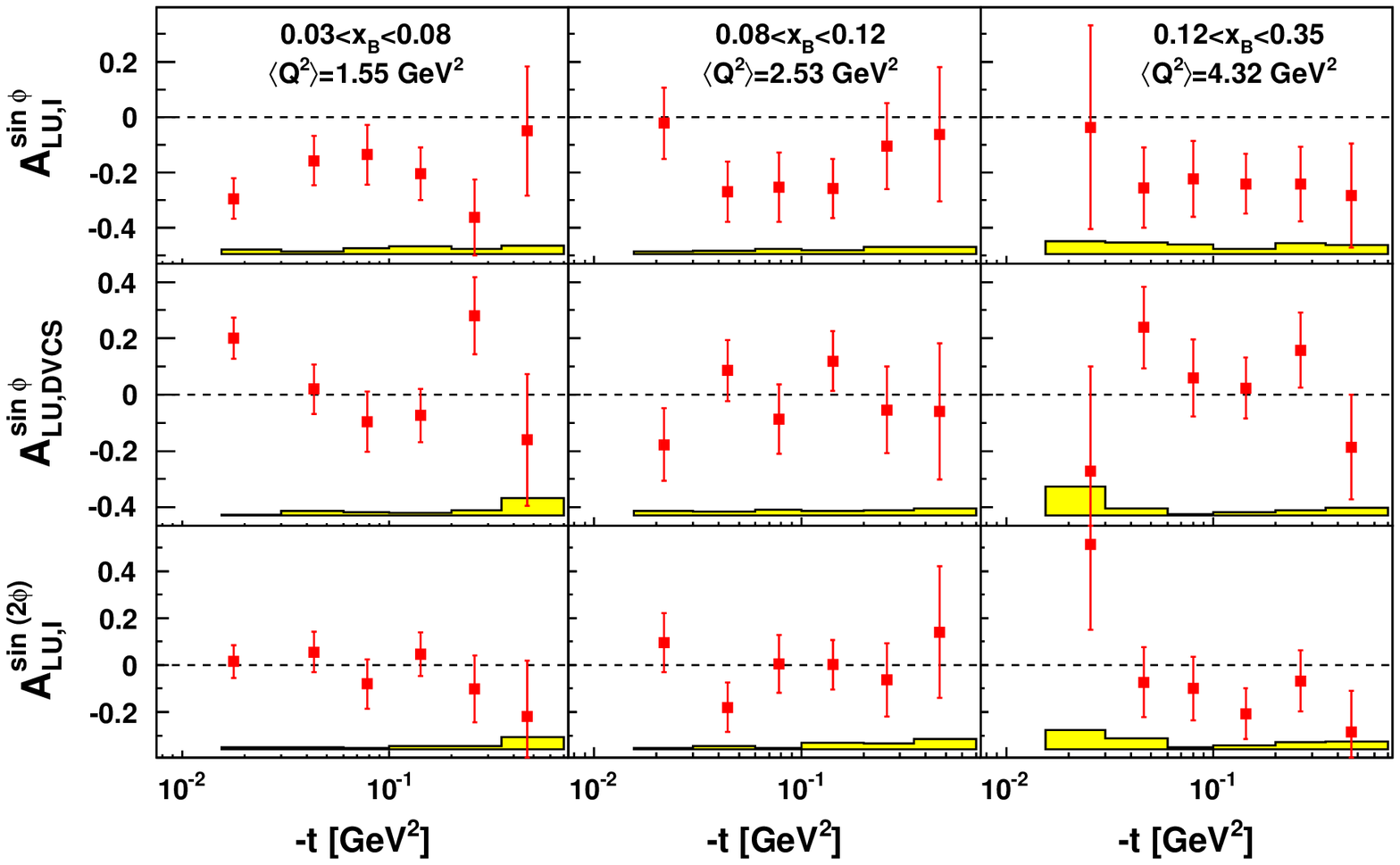}
\caption{
The first (second) row shows the $\sin \phi$ amplitude 
of the beam-helicity asymmetry $\ALUI$ ($\ALUDVCS$) sensitive to the 
interference term (squared DVCS term),
extracted from the 1996--2005 hydrogen data
as a function of $-t$ 
for three $x_{\text{B}}$ ranges.
Correspondingly, the third row shows the $\sin (2 \phi)$ amplitude of $\ALUI$.
The error bars (bands) represent the statistical (systematic) uncertainties.
Not included is a 2.8\% scale uncertainty due to 
the beam polarization measurement.}
 \label{2dim_bsa_fig}
}
as a function of $-t$ for three different ranges of $x_{\text{B}}$.
The possibly negative $\sin 2 \phi$ amplitude
for the largest $x_{\text{B}}$ bins in Fig.~\ref{bsa_fig} is found to be independent of $t$
in the lower right panel of Fig.~\ref{2dim_bsa_fig}.

The $\cos (n \phi)$ amplitudes ($n = 0\text{--}3$) of the beam charge asymmetry
are shown in Fig.~\ref{bca}.
\FIGURE{
\includegraphics[width=\columnwidth]{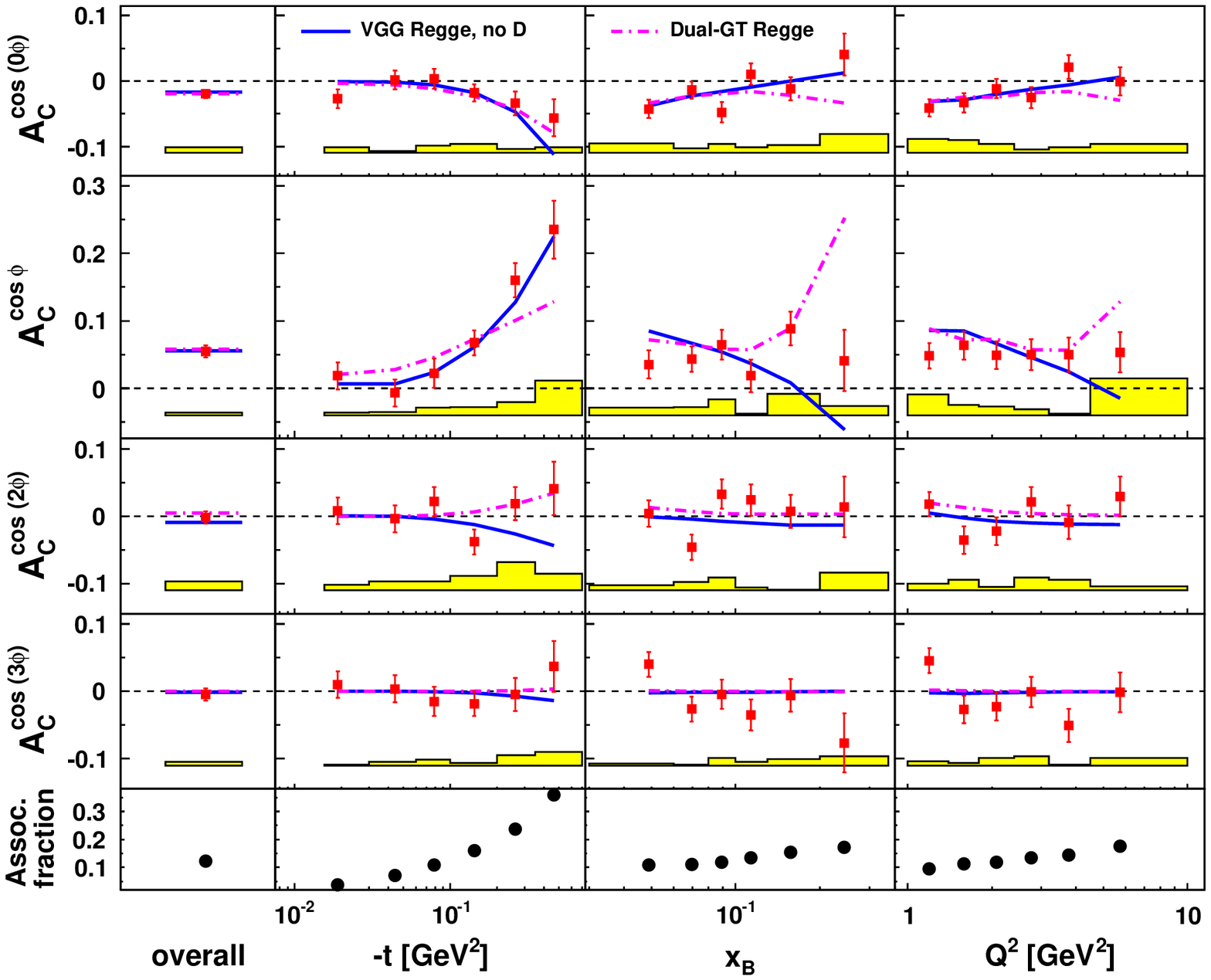}
\caption{
The $\cos (n \phi)$ amplitude ($n = 0\text{--}3$) of the beam-charge asymmetry $\AC$,
extracted from the 1996--2005 hydrogen data in the entire experimental acceptance, and 
as a function of $-t$, $x_{\text{B}}$, and $Q^2$.
The error bars (bands) represent the statistical (systematic) uncertainties.
The theoretical calculations are based on the models that are unable to 
describe the data in Fig.~\ref{bsa_fig}.
For the VGG model the parameter settings $b_{\mathrm{val}} = \infty$ and
$b_{\text{sea}} = 1$ are used and the contribution from the D--term is set to zero.
The bottom row shows the fractional contribution of associated BH production as obtained from a MC simulation.}
 \label{bca}
}
The $\cos (0\phi)$  and the $\cos \phi$ amplitudes, i.e., the amplitudes related to
twist-2 GPDs, are zero at small values of $-t$ and become
non-zero with increasing values of $-t$, with opposite sign
and smaller magnitude for $\cos (0\phi)$ as expected from Eq.~\ref{r}.
It is interesting to note that $\AC^{\cos \phi }$ and $\ALUI^{\sin \phi}$
show a fundamentally different dependence on $-t$, despite relating
to the real and imaginary parts of the twist-2 helicity-conserving 
DVCS amplitude via the same factor $k \propto \sqrt{-t}/Q$ in
Eq.~\ref{c1I} and Eq.~\ref{s1I}, respectively.
The $\cos \phi$ amplitude does not exhibit any kinematic dependence on $x_{\text{B}}$ or $Q^2$. 
It is in agreement with the earlier HERMES measurements based on subsamples 
of the data used in this analysis~\cite{Air06,Air08}. 
The $\cos (2 \phi)$ amplitude, which is related to twist-3 GPDs, is suppressed as
expected and found to be compatible with zero.
Also, the $\cos (3 \phi)$ amplitude, which is related to
gluon helicity-flip GPDs, is found to be consistent with zero. 
No striking additional features are observed in Fig.~\ref{2dim_bca_fig} 
where the $\cos (n \phi)$ amplitudes are shown  
as a function of $-t$ for three distinct $x_{\text{B}}$ ranges.
\FIGURE{
\includegraphics[width=\columnwidth]{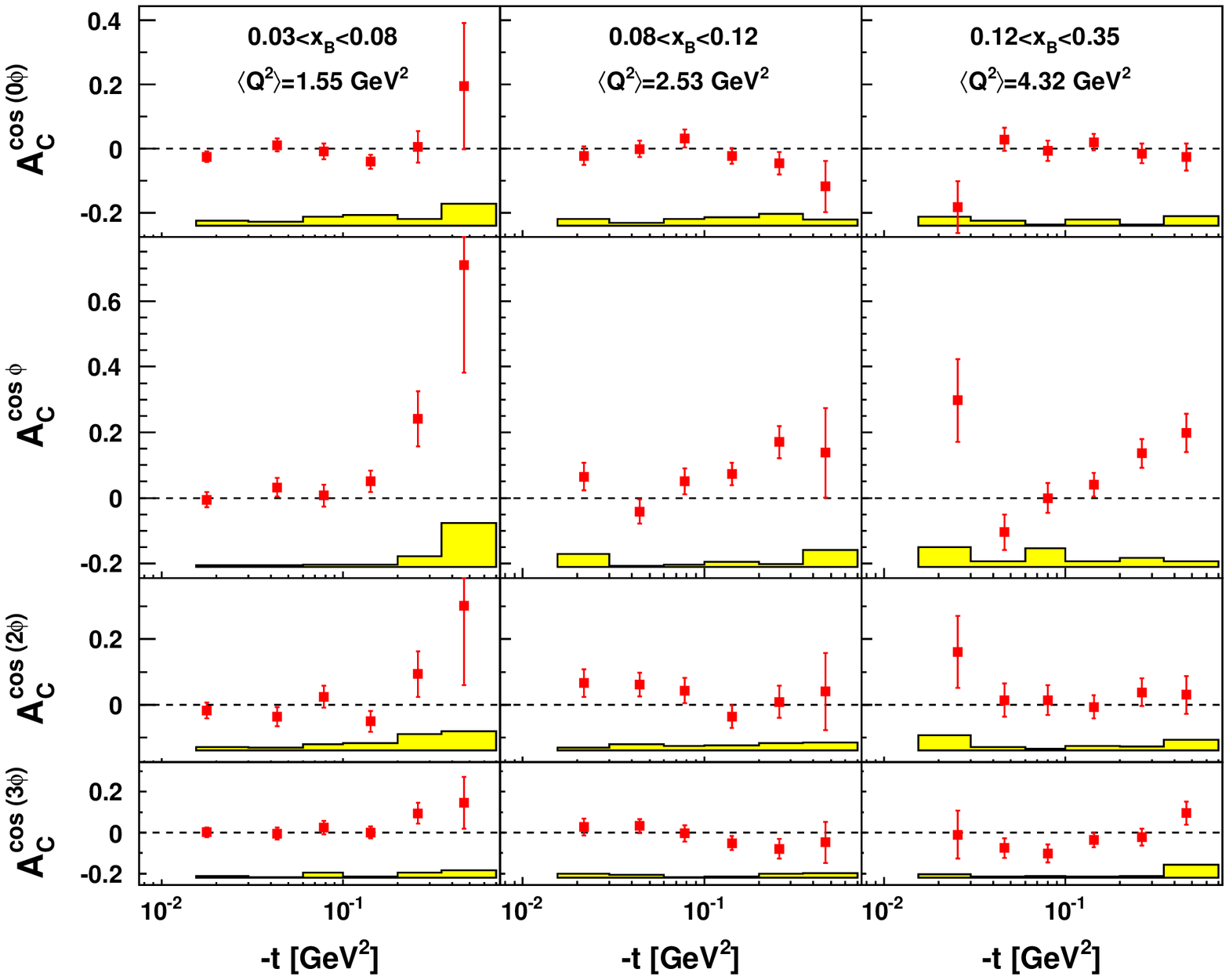}
\caption{
The $\cos (n \phi)$ amplitude ($n = 0\text{--}3$) of the beam-charge asymmetry $\AC$,
extracted from the 1996--2005 hydrogen data
as a function of $-t$ 
for three $x_{\text{B}}$ ranges.
The error bars (bands) represent the statistical (systematic) uncertainties.}
 \label{2dim_bca_fig}
}

The theoretical calculations shown in Fig.~\ref{bca} are based on either the Dual-GT 
or the VGG model. 
For the VGG model the parameter settings $b_{\mathrm{val}} = \infty$ and
$b_{\text{sea}} = 1$ are used
and the contribution from the D--term is set to zero, 
as only this set of parameters yields a good description of the BCA data~\cite{Air06, Air08}.
Note that the same set, in particular the setting $b_{\text{sea}} = 1$,  
leads to amplitudes with the largest magnitude among those represented in the bands 
in the top row of Fig.~\ref{bsa_fig}, 
i.e., it clearly does not describe the data related to
the imaginary part of the DVCS amplitude.
It appears that additional degrees of freedom in the calculation of the BCA, 
such as the value assigned to the D--term, allow the VGG model to be tuned to
resemble the BCA data.
Similarly, the Dual-GT model does not describe the data in Fig.~\ref{bsa_fig} but 
is in reasonable agreement with the BCA data shown in Fig.~\ref{bca}.
(The sudden increase of the $\cos \phi$ amplitude predicted by this model in the
highest $x_B$ and $Q^2$ bins might be due to the fact that this model is designed for 
small and medium values of $x_B$ up to 0.2.)
While the increase \{decrease\} of the $\cos \phi$ \{$\cos (0 \phi)$\} amplitude with $-t$ 
is well reproduced within these
models, the contribution of associated processes not included in these models
is expected to also increase with $-t$ as shown in the bottom row.

\section{Summary}
Previously unmeasured charge-difference and charge-averaged beam-helicity 
asymmetries in hard electroproduction of real photons from an unpolarized proton 
target are extracted from data taken with electron and positron beams.
The $\sin \phi$ amplitudes of these beam-helicity asymmetries are 
sensitive to the interference term (twist-2) and to 
the squared DVCS term (twist-3), respectively, 
whereas earlier measured beam-helicity asymmetries with a single beam-charge are 
sensitive to only their linear combination.
In addition, the most precise determination of the beam-charge asymmetry is presented, 
which provides access to the real part of the DVCS amplitude.
The GPD models presented are not able to describe 
the $\sin \phi$ amplitude sensitive to the interference term, 
while 
they can be adjusted to resemble the results on the beam-charge asymmetry, 
presumably because the model calculations have additional degrees of freedom in the latter case.
The amplitudes related to higher-twist or gluon helicity-flip GPDs 
are found to be compatible with zero.
The results presented on these charge-decomposed beam-helicity asymmetries and on the high-precision
beam-charge asymmetry
have the potential to
considerably constrain 
the GPD $H$ when used in comparison with future GPD models or as input to global fits.

\acknowledgments
We gratefully acknowledge the \desy\ management for its support and the staff
at \desy\ and the collaborating institutions for their significant effort.
This work was supported by the FWO-Flanders and IWT, Belgium;
the Natural Sciences and Engineering Research Council of Canada;
the National Natural Science Foundation of China;
the Alexander von Humboldt Stiftung;
the German Bundesministerium f\"ur Bildung und Forschung (BMBF);
the Deutsche Forschungsgemeinschaft (DFG);
the Italian Istituto Nazionale di Fisica Nucleare (INFN);
the MEXT, JSPS, and G-COE of Japan;
the Dutch Foundation for Fundamenteel Onderzoek der Materie (FOM);
the U.K.~Engineering and Physical Sciences Research Council, 
the Science and Technology Facilities Council,
and the Scottish Universities Physics Alliance;
the U.S.~Department of Energy (DOE) and the National Science Foundation (NSF);
the Russian Academy of Science and the Russian Federal Agency for 
Science and Innovations;
the Ministry of Economy and the Ministry of Education and Science of 
Armenia;
and the European Community-Research Infrastructure Activity under the
FP6 ``Structuring the European Research Area'' program
(HadronPhysics, contract number RII3-CT-2004-506078).


\newpage

\appendix
\section{Tables of Results}
\TABLE{
\tiny
\begin{tabular}{|c|c|c|c|c|r|r|r|} \hline
\multicolumn{2}{|c|}{kinematic bin} & $\langle-t\rangle$ & $\langle x_{\text{B}}\rangle$ & $\langle Q^2 \rangle $ &  \multicolumn{1}{c|}{$A_{\text{LU,I}}^{\sin \phi}$} &  \multicolumn{1}{c|}{$A_{\text{LU,DVCS}}^{\sin \phi }$} &  \multicolumn{1}{c|}{$A_{\text{LU,I}}^{\sin (2\phi) }$} \\ 
\multicolumn{2}{|c|}{} &  $[\text{GeV}^2]$ & & $[\text{GeV}^2]$ & $\pm \delta \text{(stat.)} \pm \delta \text{(syst.)}$& $\pm \delta \text{(stat.)} \pm \delta \text{(syst.)}$ & $\pm \delta \text{(stat.)} \pm \delta \text{(syst.)}$ \\ \hline \hline
\multicolumn{2}{|c|}{overall}	& $-0.12$	&	$0.10$	&	$2.46$	&	$-0.224\pm0.028\pm0.020$	&	$~0.043\pm0.028\pm0.004$	&	$-0.035\pm0.028\pm0.002$ \\ \hline
\multirow{6}{*}{\rotatebox{90}{\mbox{$-t [\text{GeV}^2]$}}} & $0.00-0.03$	&	$-0.02$	&	$0.07$	&	$1.71$	&	$-0.225\pm0.062\pm0.010$	&	$0.095\pm0.062\pm0.007$	&	$0.057\pm0.061\pm0.006$ \\
& $0.03-0.06$	&	$-0.04$	&	$0.09$	&	$2.22$	&	$-0.231\pm0.063\pm0.016$	&	$~0.091\pm0.062\pm0.010$	&	$-0.039\pm0.061\pm0.012$ \\
& $0.06-0.10$	&	$-0.08$	&	$0.10$	&	$2.44$	&	$-0.193\pm0.069\pm0.009$	&	$-0.051\pm0.069\pm0.011$	&	$-0.063\pm0.068\pm0.006$ \\
& $0.10-0.20$	&	$-0.14$	&	$0.11$	&	$2.72$	&	$-0.249\pm0.059\pm0.013$	&	$~0.020\pm0.058\pm0.008$	&	$-0.041\pm0.058\pm0.012$ \\
& $0.20-0.35$	&	$-0.26$	&	$0.12$	&	$3.13$	&	$-0.256\pm0.080\pm0.013$	&	$~0.126\pm0.079\pm0.015$	&	$-0.065\pm0.080\pm0.013$ \\
& $0.35-0.70$	&	$-0.46$	&	$0.12$	&	$3.63$	&	$-0.158\pm0.115\pm0.013$	&	$-0.101\pm0.114\pm0.010$	&	$-0.201\pm0.116\pm0.025$ \\ \hline
\multirow{6}{*}{\rotatebox{90}{\mbox{$x_{\text{B}}$}}} & $0.03-0.06$	&	$-0.10$	&	$0.05$	&	$1.34$	&	$-0.248\pm0.060\pm0.067$	&	$0.087\pm0.059\pm0.011$	&	$-0.028\pm0.059\pm0.008$ \\
& $0.06-0.08$	&	$-0.09$	&	$0.07$	&	$1.78$	&	$-0.191\pm0.059\pm0.034$	&	$~0.023\pm0.058\pm0.007$	&	$~0.011\pm0.057\pm0.003$ \\
& $0.08-0.10$	&	$-0.11$	&	$0.09$	&	$2.30$	&	$-0.215\pm0.069\pm0.018$	&	$~0.026\pm0.069\pm0.014$	&	$~0.046\pm0.068\pm0.018$ \\
& $0.10-0.13$	&	$-0.12$	&	$0.11$	&	$2.92$	&	$-0.248\pm0.071\pm0.032$	&	$~0.033\pm0.071\pm0.016$	&	$-0.039\pm0.072\pm0.010$ \\
& $0.13-0.20$	&	$-0.16$	&	$0.16$	&	$4.04$	&	$-0.244\pm0.077\pm0.023$	&	$~0.066\pm0.077\pm0.016$	&	$-0.229\pm0.076\pm0.012$ \\
& $0.20-0.35$	&	$-0.23$	&	$0.24$	&	$6.11$	&	$-0.040\pm0.139\pm0.005$	&	$-0.196\pm0.137\pm0.048$	&	$-0.138\pm0.132\pm0.022$ \\ \hline
\multirow{6}{*}{\rotatebox{90}{\mbox{$Q^2 [\text{GeV}^2]$}}} & $1.0-1.4$	&	$-0.08$	&	$0.05$	&	$1.20$	&	$-0.247\pm0.059\pm0.055$	&	$0.078\pm0.059\pm0.010$	&	$-0.034\pm0.058\pm0.005$ \\
 & $1.4-1.8$	&	$-0.10$	&	$0.07$	&	$1.59$	&	$-0.151\pm0.067\pm0.042$	&	$~0.034\pm0.067\pm0.016$	&	$~0.042\pm0.066\pm0.004$ \\
 & $1.8-2.4$	&	$-0.11$	&	$0.08$	&	$2.08$	&	$-0.230\pm0.064\pm0.031$	&	$~0.052\pm0.064\pm0.013$	&	$-0.047\pm0.062\pm0.009$ \\
 & $2.4-3.2$	&	$-0.13$	&	$0.10$	&	$2.77$	&	$-0.425\pm0.068\pm0.041$	&	$~0.212\pm0.068\pm0.015$	&	$~0.034\pm0.070\pm0.011$ \\
 & $3.2-4.5$	&	$-0.15$	&	$0.13$	&	$3.76$	&	$-0.103\pm0.076\pm0.040$	&	$-0.097\pm0.075\pm0.012$	&	$-0.116\pm0.076\pm0.003$ \\
 & $4.5-10.$	&	$-0.22$	&	$0.20$	&	$5.75$	&	$-0.129\pm0.094\pm0.008$	&	$-0.125\pm0.093\pm0.013$	&	$-0.161\pm0.092\pm0.009$ \\ \hline
\end{tabular}
\caption{
Bin sizes, average kinematic values and results of the asymmetry amplitudes 
presented in Fig.~\ref{bsa_fig}.}
\label{tab:bsa1d_res}
}

\TABLE{
\tiny
\begin{tabular}{|cc|c|c|c|c|r|r|r|} \hline
\multicolumn{3}{|c|}{kinematic bin} & $\langle-t\rangle$ & $\langle x_{\text{B}}\rangle$ & $\langle Q^2 \rangle $ & \multicolumn{1}{c|}{$A_{\text{LU,I}}^{\sin \phi}$} & \multicolumn{1}{c|}{$A_{\text{LU,DVCS}}^{\sin \phi}$} & \multicolumn{1}{c|}{$A_{\text{LU,I}}^{\sin (2\phi)}$} \\ 
\multicolumn{3}{|c|}{} &  $[\text{GeV}^2]$ & & $[\text{GeV}^2]$ & $\pm \delta \text{(stat.)} \pm \delta \text{(syst.)}$& $\pm \delta \text{(stat.)} \pm \delta \text{(syst.)}$ & $\pm \delta \text{(stat.)} \pm \delta \text{(syst.)}$ \\ \hline \hline
\multirow{6}{*}{\rotatebox{90}{\mbox{$-t [\text{GeV}^2]$}}} & \multirow{6}{*}{\rotatebox{90}{\mbox{$ 0.03 < x_{\text{B}} < 0.08$}}} & $0.00-0.03$ 	&	$-0.02$	&	$0.06$	&	$1.47$	&	$-0.295\pm0.073\pm0.016$	&	$0.201\pm0.072\pm0.004$	&	$0.015\pm0.071\pm0.008$ \\
&  & $0.03-0.06$ &	$-0.04$	&	$0.06$	&	$1.56$	&	$-0.158\pm0.089\pm0.010$	&	$0.019\pm0.089\pm0.017$	&	$0.056\pm0.085\pm0.010$ \\
&  & $0.06-0.10$ &  	$-0.08$	&	$0.06$	&	$1.55$	&	$-0.136\pm0.108\pm0.022$	&	$-0.096\pm0.107\pm0.011$	&	$-0.081\pm0.106\pm0.006$ \\
&  & $0.10-0.20$ &  	$-0.14$	&	$0.06$	&	$1.57$	&	$-0.206\pm0.095\pm0.027$	&	$-0.074\pm0.095\pm0.010$	&	$0.046\pm0.093\pm0.014$ \\
&  & $0.20-0.35$ &  	$-0.26$	&	$0.06$	&	$1.69$	&	$-0.362\pm0.137\pm0.019$	&	$0.280\pm0.137\pm0.018$	&	$-0.101\pm0.143\pm0.015$ \\
&  & $0.35-0.70$ &  	$-0.46$	&	$0.05$	&	$1.78$	&	$-0.050\pm0.234\pm0.031$	&	$-0.161\pm0.234\pm0.063$	&	$-0.218\pm0.239\pm0.052$ \\ \hline
\multirow{6}{*}{\rotatebox{90}{\mbox{$-t [\text{GeV}^2]$}}} & \multirow{6}{*}{\rotatebox{90}{\mbox{$ 0.08 < x_{\text{B}} < 0.12$ }}}& $0.00-0.03$ 	&	$-0.02$	&	$0.09$	&	$2.32$	&	$-0.022\pm0.129\pm0.010$	&	$-0.177\pm0.128\pm0.016$	&	$0.095\pm0.126\pm0.007$ \\
 & & $0.03-0.06$ &  	$-0.04$	&	$0.10$	&	$2.50$	&	$-0.269\pm0.110\pm0.011$	&	$0.086\pm0.109\pm0.013$	&	$-0.180\pm0.105\pm0.016$ \\
 & & $0.06-0.10$ &  	$-0.08$	&	$0.10$	&	$2.43$	&	$-0.254\pm0.125\pm0.018$	&	$-0.088\pm0.123\pm0.021$	&	$0.005\pm0.124\pm0.007$ \\
 & & $0.10-0.20$ &  	$-0.14$	&	$0.10$	&	$2.51$	&	$-0.258\pm0.107\pm0.015$	&	$0.119\pm0.106\pm0.018$	&	$0.001\pm0.106\pm0.028$ \\
 & & $0.20-0.35$ &  	$-0.26$	&	$0.10$	&	$2.74$	&	$-0.106\pm0.156\pm0.025$	&	$-0.054\pm0.154\pm0.019$	&	$-0.063\pm0.155\pm0.025$ \\
 & & $0.35-0.70$ &  	$-0.47$	&	$0.10$	&	$3.25$	&	$-0.062\pm0.242\pm0.026$	&	$-0.060\pm0.242\pm0.026$	&	$0.140\pm0.280\pm0.045$ \\ \hline
\multirow{6}{*}{\rotatebox{90}{\mbox{$-t [\text{GeV}^2]$}}} & \multirow{6}{*}{\rotatebox{90}{\mbox{$ 0.12 < x_{\text{B}} < 0.35$}}} & $0.00-0.03$ 	&	$-0.03$	&	$0.13$	&	$2.91$	&	$-0.037\pm0.368\pm0.046$	&	$-0.271\pm0.370\pm0.104$	&	$0.514\pm0.363\pm0.083$ \\
& & $0.03-0.06$ &  	$-0.05$	&	$0.15$	&	$3.62$	&	$-0.255\pm0.145\pm0.042$	&	$0.238\pm0.145\pm0.027$	&	$-0.073\pm0.150\pm0.048$ \\
& & $0.06-0.10$ &  	$-0.08$	&	$0.16$	&	$3.93$	&	$-0.223\pm0.137\pm0.035$	&	$0.059\pm0.137\pm0.006$	&	$-0.099\pm0.135\pm0.008$ \\
& & $0.10-0.20$ &  	$-0.14$	&	$0.17$	&	$4.30$	&	$-0.241\pm0.109\pm0.018$	&	$0.023\pm0.108\pm0.012$	&	$-0.208\pm0.108\pm0.018$ \\
& & $0.20-0.35$ &  	$-0.26$	&	$0.18$	&	$4.76$	&	$-0.242\pm0.135\pm0.039$	&	$0.158\pm0.134\pm0.018$	&	$-0.068\pm0.130\pm0.033$ \\
& & $0.35-0.70$ &  	$-0.46$	&	$0.19$	&	$5.52$	&	$-0.284\pm0.188\pm0.034$	&	$-0.187\pm0.185\pm0.028$	&	$-0.284\pm0.175\pm0.035$ \\ \hline
\end{tabular}
\caption{
Bin sizes, average kinematic values and results of the asymmetry amplitudes 
presented in
Fig.~\ref{2dim_bsa_fig}.}
\label{tab:bsa2d_res}
}

\TABLE{
\tiny
\begin{tabular}{|c|c|c|c|c|r|r|r|r|} \hline
\multicolumn{2}{|c|}{kinematic bin} & $\langle-t\rangle$ & $\langle x_{\text{B}}\rangle$ & $\langle Q^2 \rangle $ &  \multicolumn{1}{c|}{$A_{\text{C}}^{\cos(0\phi)}$} &  \multicolumn{1}{c|}{$A_{\text{C}}^{\cos \phi }$} &  \multicolumn{1}{c|}{$A_{\text{C}}^{\cos (2\phi)}$} &  \multicolumn{1}{c|}{$A_{\text{C}}^{\cos (3\phi) }$} \\ 
\multicolumn{2}{|c|}{} &  $[\text{GeV}^2]$ & & $[\text{GeV}^2]$ & $\pm \delta \text{(stat.)} \pm \delta \text{(syst.)}$& $\pm \delta \text{(stat.)} \pm \delta \text{(syst.)}$ & $\pm \delta \text{(stat.)} \pm \delta \text{(syst.)}$ & $\pm \delta \text{(stat.)} \pm \delta \text{(syst.)}$\\ \hline \hline
\multicolumn{2}{|c|}{overall}   &	$-0.12$	&	$0.10$	&	$2.46$	&	$-0.020\pm0.006\pm0.008$	&	$0.055\pm0.009\pm0.004$	&	$-0.002\pm0.009\pm0.013$	&	$-0.004\pm0.009\pm0.006$ \\ \hline
\multirow{6}{*}{\rotatebox{90}{\mbox{$-t~[\text{GeV}^2]$ }}} & $0.00-0.03$	&	$-0.02$	&	$0.07$	&	$1.71$	&	$-0.027\pm0.014\pm0.009$	&	$0.018\pm0.020\pm0.004$	&	$0.008\pm0.020\pm0.008$	&	$0.010\pm0.020\pm0.001$ \\
& $0.03-0.06$	&	$-0.04$	&	$0.09$	&	$2.22$	&	$0.001\pm0.014\pm0.003$	&	$-0.007\pm0.020\pm0.004$	&	$-0.004\pm0.020\pm0.013$	&	$0.004\pm0.020\pm0.005$ \\
& $0.06-0.10$	&	$-0.08$	&	$0.10$	&	$2.44$	&	$0.003\pm0.015\pm0.011$	&	$0.022\pm0.022\pm0.011$	&	$0.022\pm0.022\pm0.013$	&	$-0.015\pm0.022\pm0.009$ \\
& $0.10-0.20$	&	$-0.14$	&	$0.11$	&	$2.72$	&	$-0.018\pm0.013\pm0.013$	&	$0.067\pm0.018\pm0.012$	&	$-0.038\pm0.018\pm0.021$	&	$-0.019\pm0.018\pm0.004$ \\
& $0.20-0.35$	&	$-0.26$	&	$0.12$	&	$3.13$	&	$-0.034\pm0.018\pm0.006$	&	$0.160\pm0.025\pm0.019$	&	$0.018\pm0.025\pm0.042$	&	$-0.005\pm0.025\pm0.015$ \\
& $0.35-0.70$	&	$-0.46$	&	$0.12$	&	$3.63$	&	$-0.056\pm0.029\pm0.009$	&	$0.235\pm0.043\pm0.051$	&	$0.041\pm0.040\pm0.025$	&	$0.037\pm0.038\pm0.020$ \\ \hline
\multirow{6}{*}{\rotatebox{90}{\mbox{$x_{\text{B}}$}}} & $0.03-0.06$	&	$-0.10$	&	$0.05$	&	$1.34$	&	$-0.043\pm0.014\pm0.014$	&	$0.035\pm0.021\pm0.011$	&	$0.004\pm0.019\pm0.007$	&	$0.040\pm0.018\pm0.003$ \\
& $0.06-0.08$	&	$-0.09$	&	$0.07$	&	$1.78$	&	$-0.014\pm0.013\pm0.007$	&	$0.043\pm0.019\pm0.012$	&	$-0.046\pm0.019\pm0.012$	&	$-0.026\pm0.019\pm0.001$ \\
& $0.08-0.10$	&	$-0.11$	&	$0.09$	&	$2.30$	&	$-0.048\pm0.016\pm0.014$	&	$0.064\pm0.022\pm0.024$	&	$0.033\pm0.022\pm0.019$	&	$-0.005\pm0.022\pm0.011$ \\
& $0.10-0.13$	&	$-0.12$	&	$0.11$	&	$2.92$	&	$0.010\pm0.017\pm0.009$	&	$0.018\pm0.024\pm0.002$	&	$0.024\pm0.023\pm0.004$	&	$-0.035\pm0.023\pm0.005$ \\
& $0.13-0.20$	&	$-0.16$	&	$0.16$	&	$4.04$	&	$-0.012\pm0.018\pm0.012$	&	$0.088\pm0.025\pm0.032$	&	$0.007\pm0.025\pm0.001$	&	$-0.006\pm0.024\pm0.009$ \\
& $0.20-0.35$	&	$-0.23$	&	$0.24$	&	$6.11$	&	$0.040\pm0.032\pm0.029$	&	$0.041\pm0.045\pm0.014$	&	$0.014\pm0.045\pm0.026$	&	$-0.076\pm0.044\pm0.014$ \\ \hline
\multirow{6}{*}{\rotatebox{90}{\mbox{$Q^2~[\text{GeV}^2]$}}} & $1.0-1.4$	&	$-0.08$	&	$0.05$	&	$1.20$	&	$-0.041\pm0.013\pm0.021$	&	$0.048\pm0.018\pm0.031$	&	$0.018\pm0.018\pm0.010$	&	$0.046\pm0.018\pm0.006$ \\
 & $1.4-1.8$	&	$-0.10$	&	$0.07$	&	$1.59$	&	$-0.033\pm0.015\pm0.020$	&	$0.063\pm0.021\pm0.015$	&	$-0.035\pm0.021\pm0.016$	&	$-0.027\pm0.021\pm0.004$ \\
 & $1.8-2.4$	&	$-0.11$	&	$0.08$	&	$2.08$	&	$-0.012\pm0.015\pm0.014$	&	$0.049\pm0.020\pm0.013$	&	$-0.023\pm0.021\pm0.005$	&	$-0.023\pm0.020\pm0.011$ \\
 & $2.4-3.2$	&	$-0.13$	&	$0.10$	&	$2.77$	&	$-0.025\pm0.016\pm0.006$	&	$0.050\pm0.023\pm0.008$	&	$0.021\pm0.022\pm0.019$	&	$-0.001\pm0.023\pm0.014$ \\
 & $3.2-4.5$	&	$-0.15$	&	$0.13$	&	$3.76$	&	$0.021\pm0.018\pm0.009$	&	$0.050\pm0.025\pm0.002$	&	$-0.009\pm0.025\pm0.015$	&	$-0.051\pm0.025\pm0.001$ \\
 & $4.5-10.$	&	$-0.22$	&	$0.20$	&	$5.75$	&	$-0.001\pm0.021\pm0.014$	&	$0.053\pm0.030\pm0.054$	&	$0.029\pm0.030\pm0.006$	&	$-0.002\pm0.030\pm0.011$ \\ \hline
\end{tabular}
\caption{
Bin sizes, average kinematic values and results of the asymmetry amplitudes 
presented in Fig.~\ref{bca}.}
\label{tab:bca1d_res}
}

\TABLE{
\tiny
\begin{tabular}{|cc|c|c|c|c|r|r|r|r|} \hline
\multicolumn{3}{|c|}{kinematic bin} & $\langle-t\rangle$ & $\langle x_{\text{B}}\rangle$ & $\langle Q^2 \rangle $ & \multicolumn{1}{c|}{$A_{\text{C}}^{\cos(0\phi)}$} & \multicolumn{1}{c|}{$A_{\text{C}}^{\cos \phi}$} & \multicolumn{1}{c|}{$A_{\text{C}}^{\cos (2\phi)}$} & \multicolumn{1}{c|}{$A_{\text{C}}^{\cos (3\phi)}$} \\ 
\multicolumn{3}{|c|}{} &  $[\text{GeV}^2]$ & & $[\text{GeV}^2]$ & $\pm \delta \text{(stat.)} \pm \delta \text{(syst.)}$& $\pm \delta \text{(stat.)} \pm \delta \text{(syst.)}$ & $\pm \delta \text{(stat.)} \pm \delta \text{(syst.)}$ & $\pm \delta \text{(stat.)} \pm \delta \text{(syst.)}$\\ \hline \hline
\multirow{6}{*}{\rotatebox{90}{\mbox{$-t [\text{GeV}^2]$}}} & \multirow{6}{*}{\rotatebox{90}{\mbox{$0.03 < x_{\text{B}} < 0.08$}}} & $0.00-0.03$ & 	$-0.02$	&	$0.06$	&	$1.47$	&	$-0.026\pm0.017\pm0.016$	&	$-0.005\pm0.023\pm0.006$	&	$-0.017\pm0.023\pm0.012$	&	$0.002\pm0.023\pm0.008$ \\
&  & $0.03-0.06$ &  	$-0.04$	&	$0.06$	&	$1.56$	&	$0.011\pm0.021\pm0.012$	&	$0.032\pm0.028\pm0.006$	&	$-0.036\pm0.029\pm0.008$	&	$-0.004\pm0.029\pm0.002$ \\
&  & $0.06-0.10$ &  	$-0.08$	&	$0.06$	&	$1.55$	&	$-0.008\pm0.024\pm0.028$	&	$0.007\pm0.034\pm0.007$	&	$0.024\pm0.033\pm0.020$	&	$0.025\pm0.033\pm0.025$ \\
&  & $0.10-0.20$ &  	$-0.14$	&	$0.06$	&	$1.57$	&	$-0.041\pm0.022\pm0.034$	&	$0.050\pm0.032\pm0.006$	&	$-0.051\pm0.031\pm0.023$	&	$0.002\pm0.030\pm0.006$ \\
&  & $0.20-0.35$ &  	$-0.26$	&	$0.06$	&	$1.69$	&	$0.006\pm0.049\pm0.022$	&	$0.241\pm0.084\pm0.033$	&	$0.094\pm0.069\pm0.051$	&	$0.095\pm0.052\pm0.025$ \\
&  & $0.35-0.70$ &  	$-0.46$	&	$0.05$	&	$1.78$	&	$0.196\pm0.196\pm0.068$	&	$0.710\pm0.328\pm0.134$	&	$0.302\pm0.242\pm0.060$	&	$0.146\pm0.126\pm0.035$ \\ \hline
\multirow{6}{*}{\rotatebox{90}{\mbox{$-t [\text{GeV}^2]$}}} & \multirow{6}{*}{\rotatebox{90}{\mbox{$0.08 < x_{\text{B}} < 0.12$}}}	& $0.00-0.03$ & 	$-0.02$	&	$0.09$	&	$2.32$	&	$-0.021\pm0.029\pm0.021$	&	$0.065\pm0.042\pm0.039$	&	$0.066\pm0.041\pm0.010$	&	$0.028\pm0.040\pm0.018$ \\
&  & $0.03-0.06$ &  	$-0.04$	&	$0.10$	&	$2.50$	&	$-0.001\pm0.025\pm0.010$	&	$-0.041\pm0.037\pm0.004$	&	$0.062\pm0.036\pm0.020$	&	$0.032\pm0.035\pm0.014$ \\
&  & $0.06-0.10$ &  	$-0.08$	&	$0.10$	&	$2.43$	&	$0.033\pm0.028\pm0.021$	&	$0.050\pm0.039\pm0.006$	&	$0.043\pm0.039\pm0.014$	&	$-0.003\pm0.040\pm0.004$ \\
&  & $0.10-0.20$ &  	$-0.14$	&	$0.10$	&	$2.51$	&	$-0.022\pm0.025\pm0.026$	&	$0.073\pm0.035\pm0.016$	&	$-0.036\pm0.035\pm0.016$	&	$-0.051\pm0.034\pm0.005$ \\
&  & $0.20-0.35$ &  	$-0.26$	&	$0.10$	&	$2.74$	&	$-0.045\pm0.035\pm0.037$	&	$0.170\pm0.049\pm0.009$	&	$0.009\pm0.048\pm0.023$	&	$-0.079\pm0.049\pm0.021$ \\
&  & $0.35-0.70$ & 	$-0.47$	&	$0.10$	&	$3.25$	&	$-0.118\pm0.081\pm0.020$	&	$0.137\pm0.137\pm0.051$	&	$0.041\pm0.117\pm0.024$	&	$-0.047\pm0.101\pm0.021$ \\ \hline
\multirow{6}{*}{\rotatebox{90}{\mbox{$-t [\text{GeV}^2]$}}} & \multirow{6}{*}{\rotatebox{90}{\mbox{$0.12 < x_{\text{B}} < 0.35$}}}& $0.00-0.03$ &	$-0.03$	&	$0.13$	&	$2.91$	&	$-0.181\pm0.081\pm0.028$	&	$0.297\pm0.126\pm0.060$	&	$0.161\pm0.110\pm0.046$	&	$-0.010\pm0.118\pm0.016$ \\
&  & $0.03-0.06$ &  	$-0.05$	&	$0.15$	&	$3.62$	&	$0.029\pm0.036\pm0.016$	&	$-0.104\pm0.054\pm0.016$	&	$0.014\pm0.051\pm0.011$	&	$-0.075\pm0.049\pm0.005$ \\
&  & $0.06-0.10$ &  	$-0.08$	&	$0.16$	&	$3.93$	&	$-0.006\pm0.032\pm0.004$	&	$-0.000\pm0.045\pm0.056$	&	$0.014\pm0.046\pm0.006$	&	$-0.101\pm0.044\pm0.008$ \\
&  & $0.10-0.20$ &  	$-0.14$	&	$0.17$	&	$4.30$	&	$0.021\pm0.025\pm0.020$	&	$0.041\pm0.036\pm0.017$	&	$-0.007\pm0.035\pm0.015$	&	$-0.036\pm0.035\pm0.006$ \\
&  & $0.20-0.35$ &  	$-0.26$	&	$0.18$	&	$4.76$	&	$-0.015\pm0.031\pm0.004$	&	$0.136\pm0.043\pm0.027$	&	$0.038\pm0.042\pm0.013$	&	$-0.022\pm0.042\pm0.008$ \\
&  & $0.35-0.70$ &  	$-0.46$	&	$0.19$	&	$5.52$	&	$-0.026\pm0.042\pm0.031$	&	$0.199\pm0.059\pm0.017$	&	$0.030\pm0.057\pm0.034$	&	$0.095\pm0.057\pm0.063$ \\ \hline
\end{tabular}
\caption{
Bin sizes, average kinematic values and results of the asymmetry amplitudes 
presented in Fig.~\ref{2dim_bca_fig}.}
\label{tab:bca2d_res}
}


\end{document}